\documentclass[12pt,a4paper,dvips]{article}
\usepackage{epsfig}
\usepackage{graphicx}
\begin{document}

\vspace*{-0.cm}



\vspace*{0.5cm}

{\LARGE\bf\centerline{Studies of Boosted Decision Trees}}
{\LARGE\bf\centerline{for MiniBooNE Particle Identification}}




\vspace*{0.5cm}
{\large\centerline{Hai-Jun Yang$^{a,c,}\footnote{ E-mail address: yhj@umich.edu}$,
 Byron P. Roe$^a$, Ji Zhu$^b$}}


\vspace*{0.5cm}
{\small\em\centerline{$^a$ Department of Physics, University of Michigan, Ann Arbor, MI 48109, USA}}
{\small\em\centerline{$^b$ Department of Statistics, University of Michigan, Ann Arbor, MI 48109, USA}}
{\small\em\centerline{$^c$ Los Alamos National Laboratory, Los Alamos, NM 87545, USA}}
\begin{abstract}


Boosted decision trees are applied to particle identification in the MiniBooNE
experiment operated at Fermi National Accelerator Laboratory (Fermilab) for
neutrino oscillations. 
Numerous attempts are made to tune the boosted decision trees, to compare performance of
various boosting algorithms, and to select input variables for optimal performance.

\end{abstract}


\section{Introduction}

In High Energy Physics (HEP) experiments, people usually need to select
some events with specific interest, so called signal events, out of numerous 
background events for study. In order to increase the ratio of signal to 
background,
one needs to suppress background events while keeping high signal efficiency. 
To this end, some advanced techniques, such as AdaBoost\cite{adaboost}, 
$\epsilon$-Boost\cite{eboost}, 
$\epsilon$-LogitBoost\cite{eboost}, $\epsilon$-HingeBoost, Random Forests
\cite{randomforests} etc.,
from Statistics and Computer Sciences 
were introduced for signal and background event separation 
in the MiniBooNE experiment\cite{boone} at Fermilab.  
The MiniBooNE experiment is designed to
confirm or refute the evidence for $\nu_\mu \rightarrow \nu_e$ oscillations at
$\Delta m^2 \simeq 1 ~eV^2/c^4$ found by the LSND experiment\cite{lsnd}.
It is a crucial experiment which will imply new physics beyond the standard model
if the LSND signal is confirmed.
These techniques are tuned with 
one sample of Monte Carlo (MC) events, the training sample, 
and then tested with an independent
MC sample, the testing sample.  
Initial comparisons of these techniques
with artificial neural networks (ANN) using the MiniBooNE 
MC samples were described 
previously\cite{nima-boosting2005}.
This work indicated that
the method of boosted decision trees is superior to the ANNs for  
Particle Identification 
(PID) using the MiniBooNE MC samples. Further studies show that the boosted 
decision tree method has not only better event separation, but is also 
more stable and robust
than ANNs when using MC samples with varying input 
parameters.

The boosting algorithm is one of the most powerful learning techniques introduced
in the past decade\cite{breiman,schapire,freund,friedman}. 
The motivation for the boosting algorithm is to design a
procedure that combines many ``weak'' classifiers to achieve a final powerful 
classifier. In the present work numerous trials are made to 
tune the boosted decision trees, and 
comparisons are made for various algorithms. For a large number of discriminant
variables, several techniques are described to select a set of 
powerful input variables in order to 
obtain optimal event separation using boosted decision trees.
Furthermore, post-fitting of weights for the trained boosting trees is also 
investigated to attempt further possible improvement.

This paper is focussed on the boosting tuning. 
All results appearing in this paper
are relative numbers. They do not represent the MiniBooNE PID performance; that
performance is continually improving with further algorithm and PID study.
The description of the MiniBooNE reconstruction 
packages\cite{rfitters,sfitters},
the reconstructed variables, the overall and absolute 
performance of the boosting PID\cite{sf2,rf2},
the validation of the input variables and the boosting PID variables by comparing
various MC and real data samples\cite{hray} will be described in future articles.

\section{Decision Trees}

Boosting algorithms can be applied to any classifier,
Here they are applied to decision trees. A schematic of a 
simple decision tree is shown in Figure 1, S means signal, B means
background, terminal nodes called leaves are shown in boxes.
The key issue is to define a criterion that describes the 
goodness of separation between signal and background in the tree split.
Assume the events are weighted with each event
having weight $W_i$.  Define the purity of the sample in a node by
$$P = {\sum_sW_s\over \sum_sW_s +\sum_bW_b},$$
where $\sum_s$ is the sum over signal events and $\sum_b$ is the sum
over background events.
Note that $P(1-P)$ is 0 if the sample is pure signal or pure background.
For a given node let
$$Gini = (\sum_{i=1}^n W_i)P(1-P),$$
where $n$ is the number of events on that node.
The criterion chosen is to minimize
$$Gini_{left\ child} + Gini_{right\ child}.$$

To determine the increase in quality when a
node is split into two nodes, one maximizes
$$ Criterion = Gini_{father} - Gini_{left\ child} - Gini_{right\ child}
.$$

At the end, if a leaf has purity greater than 1/2 (or whatever is set), then
it is called a signal leaf, otherwise, a background leaf.  
Events are classified signal (have score of 1) 
if they land on a signal leaf and background (have score of -1) 
if they land on a background leaf.  The resulting
tree is a {\it decision tree}.

Decision trees have been available for some time\cite{breiman}.
They are known to be powerful but unstable, i.e., a small change in the
training sample can produce a large change in the tree and the results.
Combining many decision trees to make a ``majority vote'', as in the 
random forests method, 
can improve the stability somewhat.  However, as will be discussed in
Section 6, the performance of the
random forests method is significantly worse than 
the performance of the boosted decision tree method in which 
the weights of misclassified events are boosted for succeeding trees.

\begin{figure}
\begin{center}
\epsfig{figure=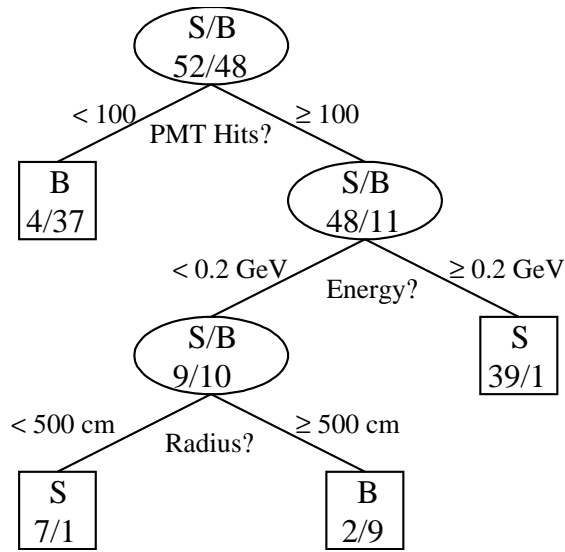,width=8cm,angle=0}
\caption{Schematic of a decision tree.}
\end{center}
\end{figure}

\section{Some Boosting Algorithms}

If there are $N$ total events in the sample, the weight of each event is 
initially taken as $1/N$. Suppose that there are $M$ trees and $m$ is the 
index of an individual tree.  Let 
\begin{itemize}
\item $x_i=$  the set of PID variables for the $i$th event.
\item $y_i =1$ if the $i$th event is a signal event
and $y_i=-1$ if the event is a background event.
\item $w_i=$ the weight of the $i$th event.  
\item $T_m(x_i)=1$ if the set of variables for the $i$th event lands that
event on a signal leaf and  $T_m(x_i)=-1$ if the set of variables for that
event lands it on a background leaf.
\item $I(y_i \ne T_m(x_i)) = 1$ if $y_i \ne T_m(x_i)$ and 0 if 
  $y_i = T_m(x_i)$.
\end{itemize}
There are several commonly used algorithms for boosting the weights of the 
misclassified events in the training sample. 
The boosting performance is quite different using various ways to update 
the event weights.

\subsection{AdaBoost}
The first boosting method is called ``AdaBoost''\cite{adaboost} or
sometimes discrete AdaBoost.
Define for the $m$th tree:
$$err_m ={\sum_{i=1}^Nw_iI(y_i \ne T_m(x_i))\over \sum_{i=1}^Nw_i}.$$
Calculate:
$$\alpha_m = \beta\times\ln((1-err_m)/err_m).$$
$\beta=1$ is the value used in the standard AdaBoost method.  
Change the weight of each event $i$, $i = 1,...,N$. 
$$w_i\rightarrow w_i\times e^{\alpha_mI(y_i\ne T_m(x_i))}.$$
Renormalize the weights.
$$w_i\rightarrow {w_i\over \sum_{i=1}^Nw_i}.$$
The score for a given event is 
$$T(x) = \sum_{m=1}^M\alpha_mT_m(x),$$
which is just the weighted sum of the scores of the individual trees.

\subsection{$\epsilon$-Boost}
A second boosting method is called
``$\epsilon$-Boost'' \cite{eboost}, or sometimes ``shrinkage''.  
After the $m$th tree, change the weight of each event $i$, $i = 1,...,N$.
$$w_i\rightarrow w_ie^{2\epsilon I(y_i\ne T_m(x_i))},$$
where $\epsilon$ is a constant of the order of 0.01.
Renormalize the weights.
$$w_i\rightarrow {w_i\over \sum_{i=1}^Nw_i}.$$
The score for a given event is 
$$T(x)  = \sum_{m=1}^M\epsilon T_m(x),$$
which is the renormalized, but unweighted, sum of the scores over individual
trees.

\subsection{$\epsilon$-LogitBoost}
A third boosting method is called ``$\epsilon$-LogitBoost''. This method is
quite similar to $\epsilon$-Boost, but the weights are updated according to:
$$w_i\rightarrow \frac{e^{-y_i T(x_i)}}{1+e^{-y_i T(x_i)}}.$$
where $T(x) = T(x) + \epsilon \times T_m(x)$  for the m$th$ tree iteration.

\subsection{$\epsilon$-HingeBoost}
A fourth boosting method is called ``$\epsilon$-HingeBoost''. Again this
method is quite similar to $\epsilon$-Boost, but here the weights are 
updated according to:
$$
w_i = 1 ~if ~y_i T(x_i)<1;~~
w_i = 0 ~if ~y_i T(x_i)\ge 1
$$
where $T(x) = T(x) + \epsilon \times T_m(x)$ for the m$th$ tree iterations.

\subsection{LogitBoost}
A fifth boosting method is called ``LogitBoost''\cite{eboost}.  
Let $y_i^*=1$ for signal
events and $y_i^*=0$ for background events.  Initial probability estimates
are set to $p(x_i) = 0.5$ for event $i$, where $x$ is the set of PID
variables.  Let:
$$z_i = {y_i^* -p(x_i)\over p(x_i)(1-p(x_i))};$$
$$w_i = p(x_i)(1-p(x_i)),$$
where $w_i$ is the weight of event $i$.  
Let $\overline{z}$ be the weighted average of $z$ over some set of events.
Instead of the Gini criterion, the splitting variable and point to divide
the events at a node into two nodes $L$ and $R$ is
determined to minimize
$$\sum_L w_i(z_i-\overline{z}_L)^2 + \sum_R w_i(z_i-\overline{z}_R)^2.$$
The output for tree $m$ is 
$T^*_m(x_i) = \overline{z}$ for the node onto which event
$i$ falls.  The total output score is:
$$T(x)=\sum_{m=1}^M {1\over 2}T^*_m(x).$$  
The probability is updated according to:
$$p(x) = {e^{T(x)}\over e^{T(x)}+e^{-T(x)}}.$$

\subsection{Gentle AdaBoost}
A sixth boosting method is called ``Gentle AdaBoost''\cite{eboost}. 
It uses same criterion as described for the LogitBoost. 
Here $z_i$ is same as $y_i$. The weights are updated according to:
$$ w_i\rightarrow w_ie^{-(2 p_m(x_i)-1)} $$
for signal events and
$$w_i\rightarrow w_ie^{+(2 p_m(x_i)-1)} $$
for background events, where $p_m(x_i)$ is the weighted purity 
of the leaf on which event $i$ falls.

\subsection{Real AdaBoost}
A seventh boosting method is called ``Real AdaBoost''\cite{eboost}.
It is similar to the discrete version of AdaBoost 
described in Section 3.1, but the weights and event scores are 
calculated in different ways. The event score for event $i$ in tree $m$
is given by:
$$T_m(x_i) = 0.5 \times \ln(p_m(x_i)/(1.-p_m(x_i)))$$
where $p_m(x_i)$ is the weighted purity of the leaf on which event $i$ falls. 
The event weights are updated according to:
$$
w_i \rightarrow w_i \times e^{(-y_i \times T_m(x_i))}
$$
and then renormalized so that the total weight is one. 
The total output score including all of the trees
is given by $T(x) = \sum_{m=1}^M T_m(x)$

\section{Tuning Parameters for the Boosted Decision Trees}


MiniBooNE MC samples from the February 2004 Baseline MC were used to
tune some parameters of the boosted decision trees. There are 88233 intrinsic
$\nu_e$ signal events and 162657 $\nu_\mu$ background events. 
20000 signal events and 30000 background events were selected randomly for 
the training sample and the rest of the events were the test sample. 
The number of input variables for boosting training is 52. 
The relative ratio is defined as the 
background efficiency divided by the corresponding signal efficiency and 
rescaled by a constant value.

\begin{figure}
\epsfig{figure=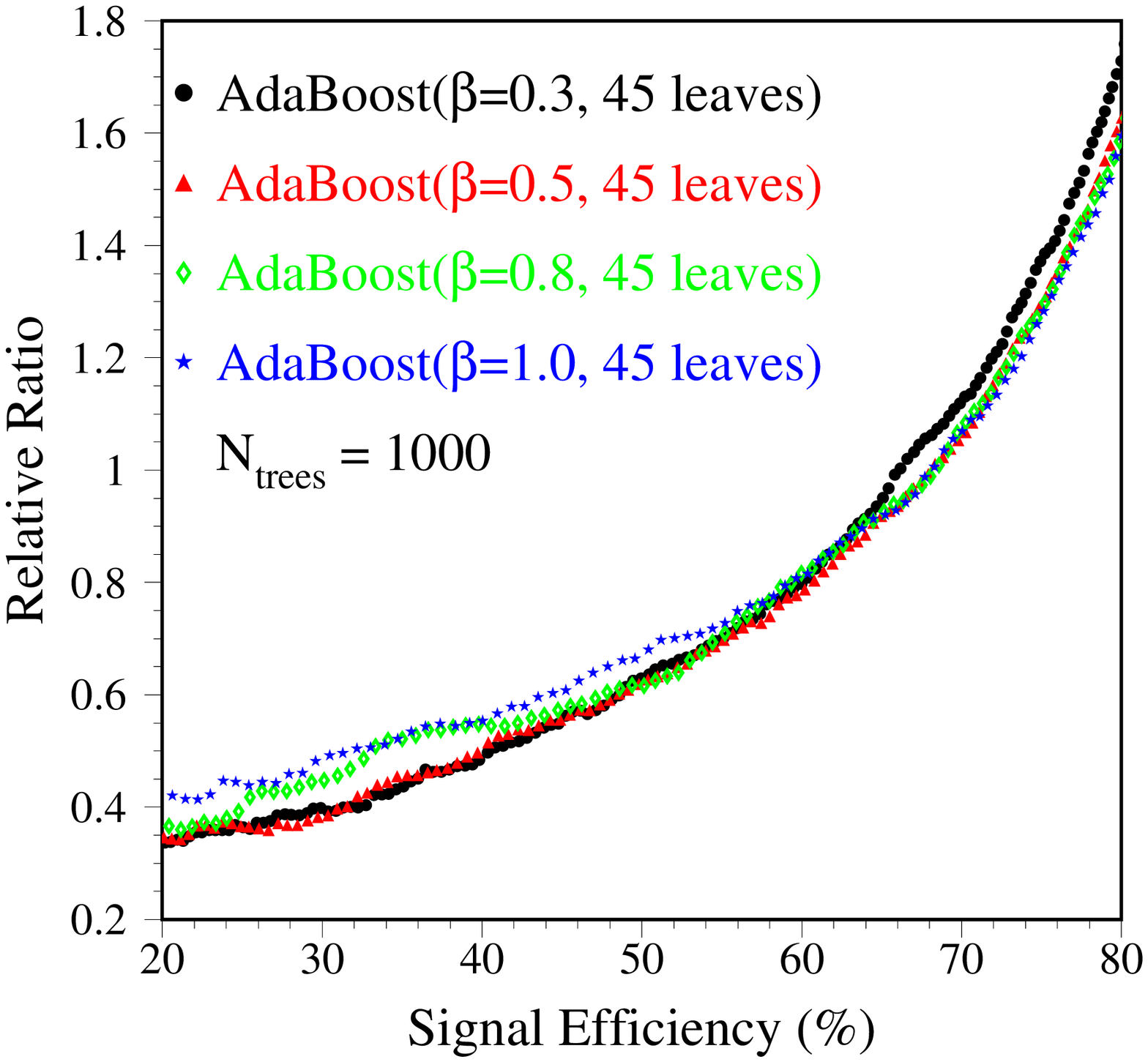,width=7.5cm}
\epsfig{figure=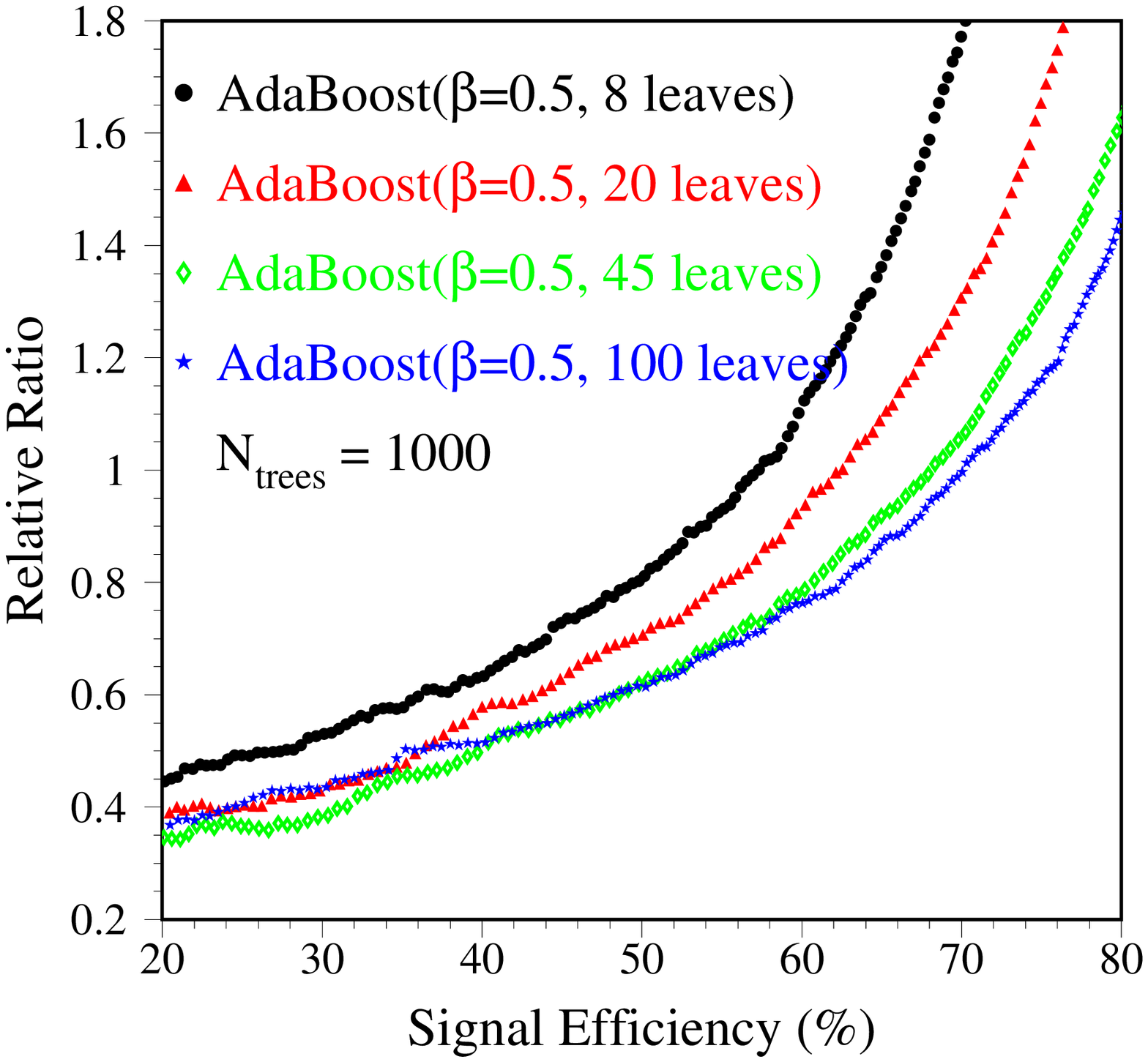,width=7.5cm}
\caption{Tuning $\beta$ (left) and decision tree size (right) using AdaBoost.}
\end{figure}

The left plot of Figure 2 shows the relative ratio versus the signal efficiency
for AdaBoost with 45 leaves per decision tree and various $\beta$ values for 1000
tree iterations. The boosting performances slightly depend on the $\beta$ values.
AdaBoost with $\beta=1$ works slightly better in the 
high signal efficiency region
(Eff $>$ 65\%) but worse in the low signal efficiency region (Eff $<$ 60\%) 
than AdaBoost 
with smaller $\beta$ values, 0.8, 0.5 or 0.3. To balance the overall performance,
$\beta=0.5$ is selected to replace the standard value 1 for the AdaBoost training. 

The right plot of Figure 2 shows the relative ratio versus the signal efficiency
for AdaBoost with $\beta=0.5$ and 1000 tree iterations for various decision tree
sizes ranging from 8 leaves to 100 leaves. AdaBoost with a large tree
size worked significantly better than AdaBoost 
with a small tree size, 8 leaves; the latter number has been 
recommended in some statistics literature\cite{leaf01}.  
Typically, it takes more tree iterations
for the smaller tree size to reach optimal performance. For this application,
even with more tree iterations (10000 trees), results from boosting with small tree size 
(8 leaves) are still significantly worse ($\sim$10\%-20\%) than results obtained 
with large tree size (45 leaves). Here, 45 leaves per decision tree 
is selected
(this number is quite close to the number of input variables, 52, for the boosting training.) 

How many decision trees are sufficient? It depends on the MC samples for boosting
training and testing.  For the given set of boosting parameters selected above, we 
ran boosting with 1000 tree iterations. The left plot of Figure 3 shows
the relative ratio versus the signal efficiency for AdaBoost with 
tree iterations
of 100, 200, 500, 800 and 1000, respectively. The boosting performance becomes better
with more tree iterations. The right plot of Figure 3 shows the
relative ratio versus the number of decision trees for  
signal efficiencies of 
50\%, 60\% and  70\% which cover the regions 
of greatest interest for the MiniBooNE experiment. 
Typically, the boosting performance for low signal efficiencies converges after few 
hundred tree iterations and is then stable. For high signal efficiency, 
boosting performance continues to  improve as the number of decision trees
is increased. 
For these particular MC samples, the boosting performance is close to optimal 
after 1000 tree iterations.  
For the sake of comparison, the AdaBoost performance of the boosting training MC
samples is also shown in the right plot of Figure 3. The relative ratios drop 
quickly down to zero (zero means no background events left after selection 
for a given signal efficiency) within 100 tree iterations for 50\%-70\% signal efficiencies.
The AdaBoost outputs for the training MC sample
and for the testing MC sample for $N_{tree} = $ 1, 100, 500 and 1000 are shown in 
Figure 4. The signal and background separation for the training sample
becomes better as the number of tree iterations increases. 
The signal and background events are
completely distinguished after about 500 tree iterations. For the testing samples, however,
the signal and background separations are quite stable after a few hundred tree iterations.
The corresponding relative ratios are stable for given signal efficiencies as shown in 
right plot of Figure 3.

\begin{figure}
\epsfig{figure=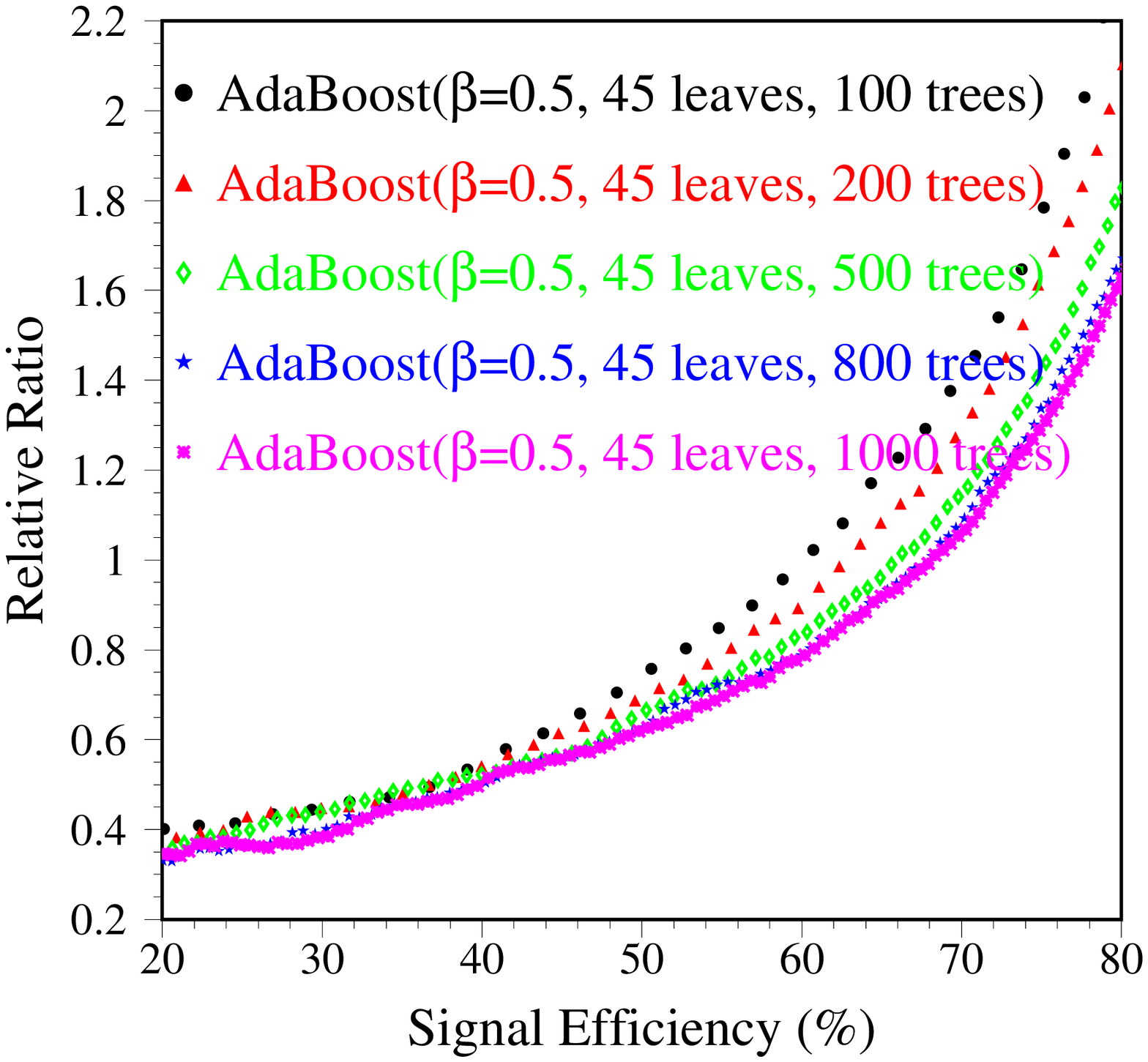,width=7.5cm}
\epsfig{figure=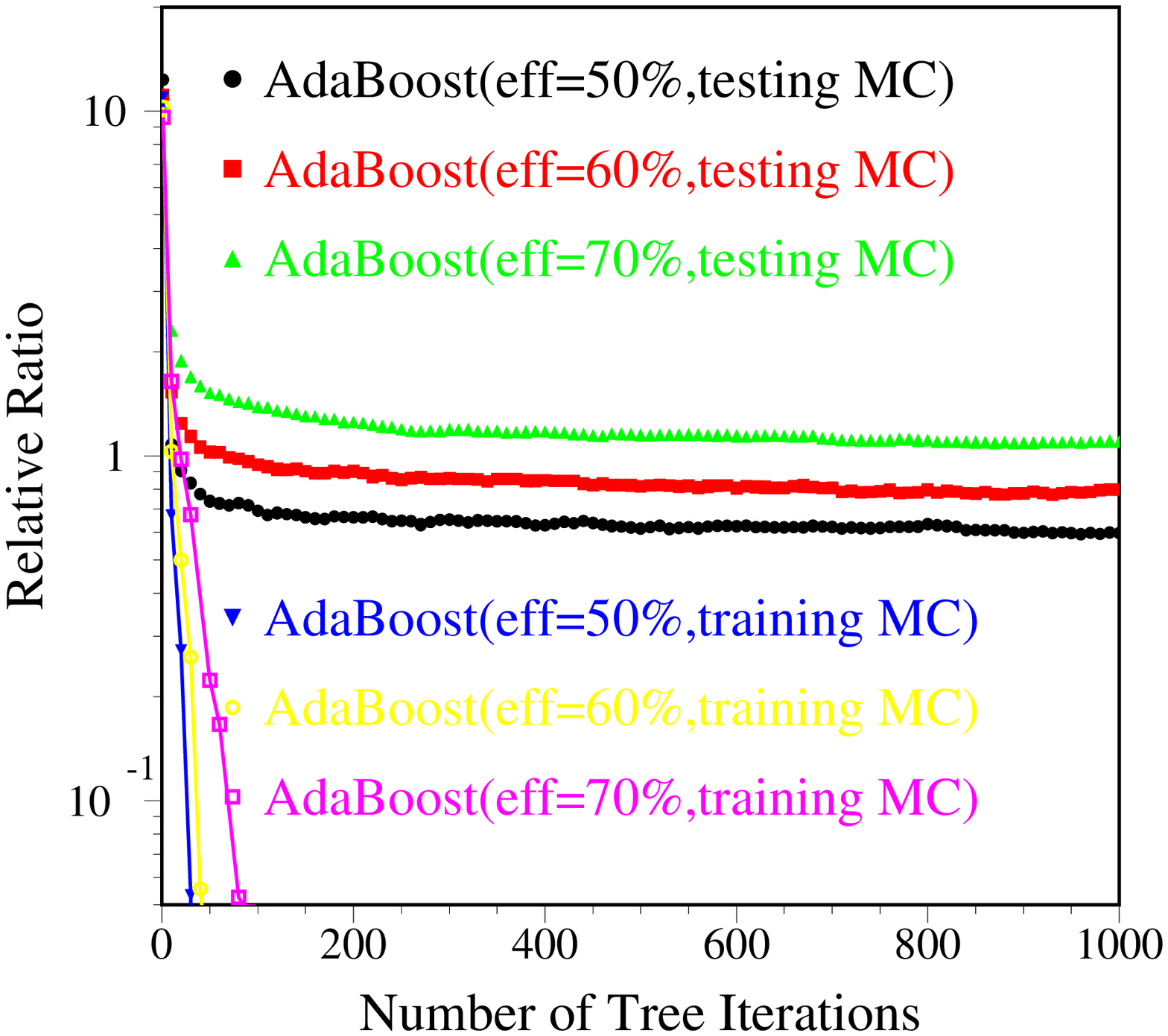,width=7.5cm}
\caption{Tuning decision tree iterations using AdaBoost (left) and 
the relative ratio versus the number of tree iterations for signal
efficiencies ranging from 50\% to 70\% (right). $\beta=0.5$ and there are 
45 leaves per decision tree.}
\end{figure}

\begin{figure}
\epsfig{figure=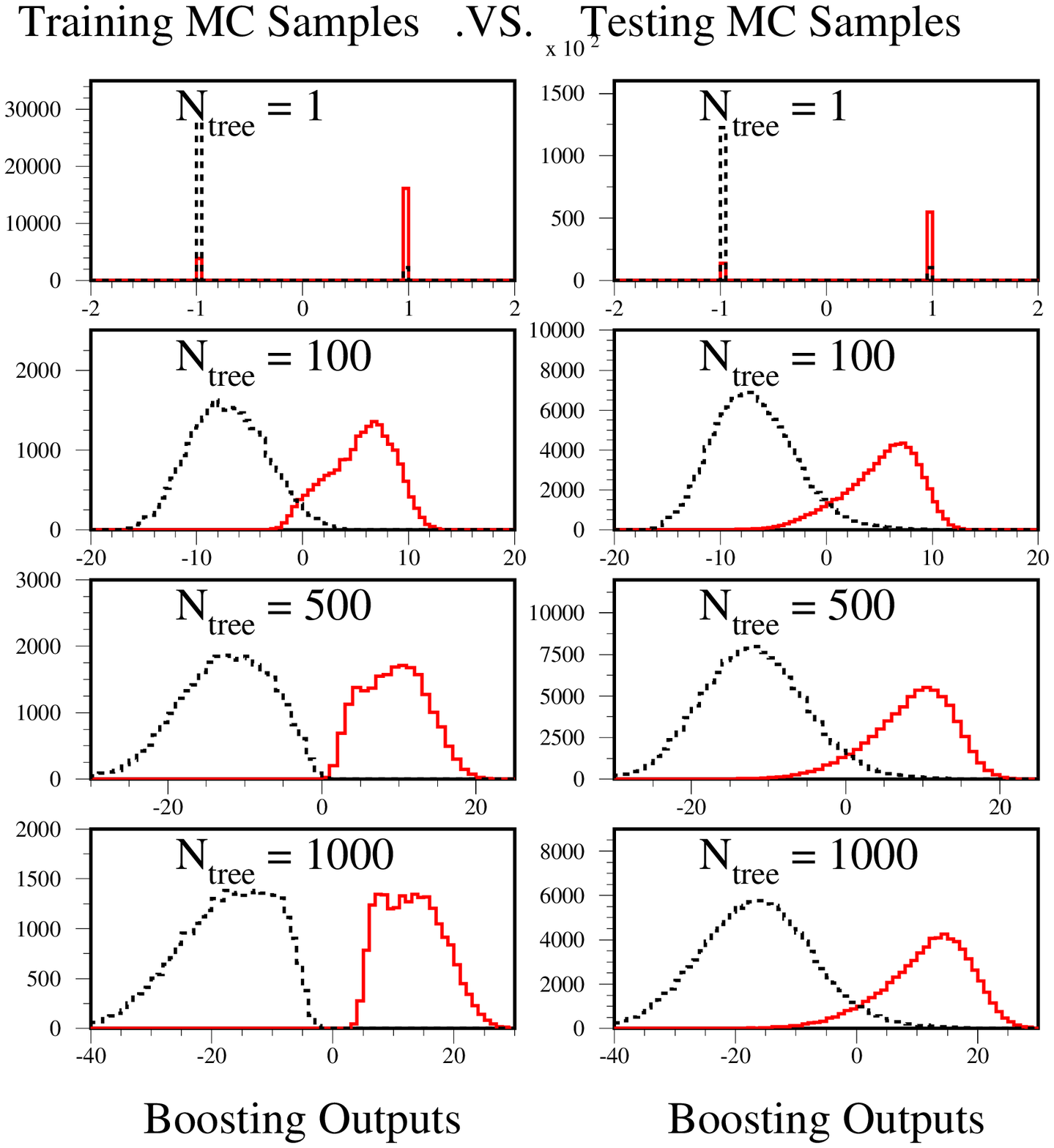,width=15cm}
\caption{The AdaBoost outputs for training MC samples (plots in left column)
and for testing MC samples (plots in right column) for $N_{tree} = $ 1, 100, 500 and 1000,
respectively. Dotted histograms represent background events and solid histograms represent signal events.
$\beta=0.5$ and there are 45 leaves per decision tree.
}
\end{figure}

The tuning parameter for $\epsilon$-Boost is $\epsilon$. The left plot of 
Figure 5 shows the relative ratio versus the signal efficiency for $\epsilon$-Boost
with $\epsilon$ values of 0.005, 0.01, 0.02, 0.04, respectively. 
$\epsilon$-Boost with fairly large $\epsilon$ values for 45 leaves per decision 
tree and 1000 tree iterations has better performance for the 
high signal efficiency 
region (Eff $>$ 50\%). The results from AdaBoost with $\beta$=0.5 are comparable to
those from $\epsilon$-Boost. $\epsilon$-Boost with $\epsilon > $ 0.01 works slightly 
better because $\epsilon$-Boost converges more quickly with larger $\epsilon$ values.
However, with more tree iterations, the final performances
for different $\epsilon$ values are very comparable. Here $\epsilon$ = 0.01 is chosen for
further comparisons.

The right plot of Figure 5 shows the relative ratio versus the signal efficiency for 
AdaBoost and $\epsilon$-Boost using two different ways to split tree nodes.
One way is to maximize the criterion based on the Gini index to select the next tree
split, the other way is to split the left tree node first. 
For AdaBoost, the performance for the ``left node first'' method 
gets worse for signal efficiency less than about 65\%. 
At about the same signal efficiency, 
the performance  for the two $\epsilon$-Boosts 
are quite comparable and are comparable with AdaBoost based on the Gini index. 
However, the $\epsilon$-Boost method based on the Gini index
becomes worse than the others for high signal efficiency. 

\begin{figure}
\epsfig{figure=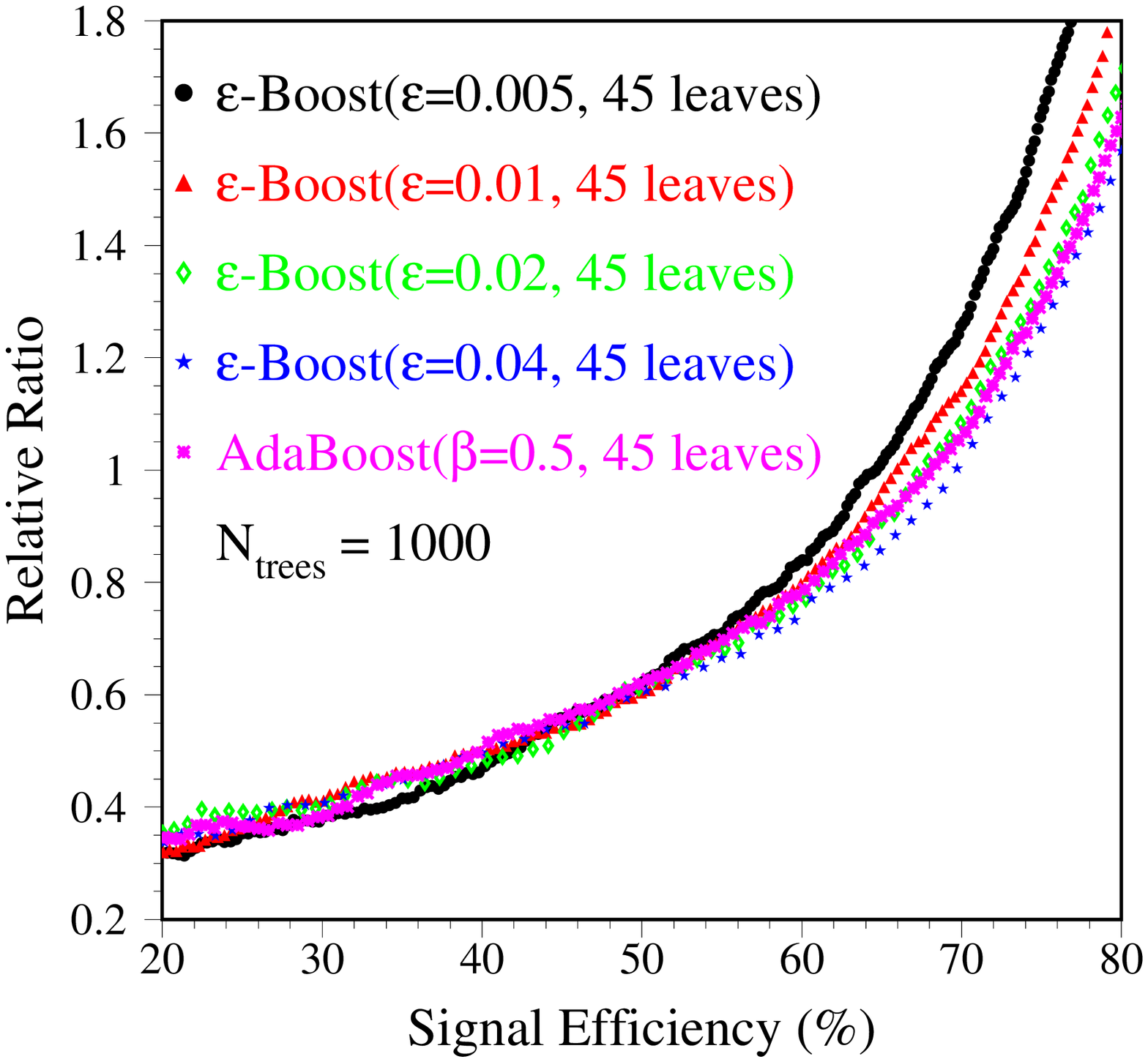,width=7.5cm}
\epsfig{figure=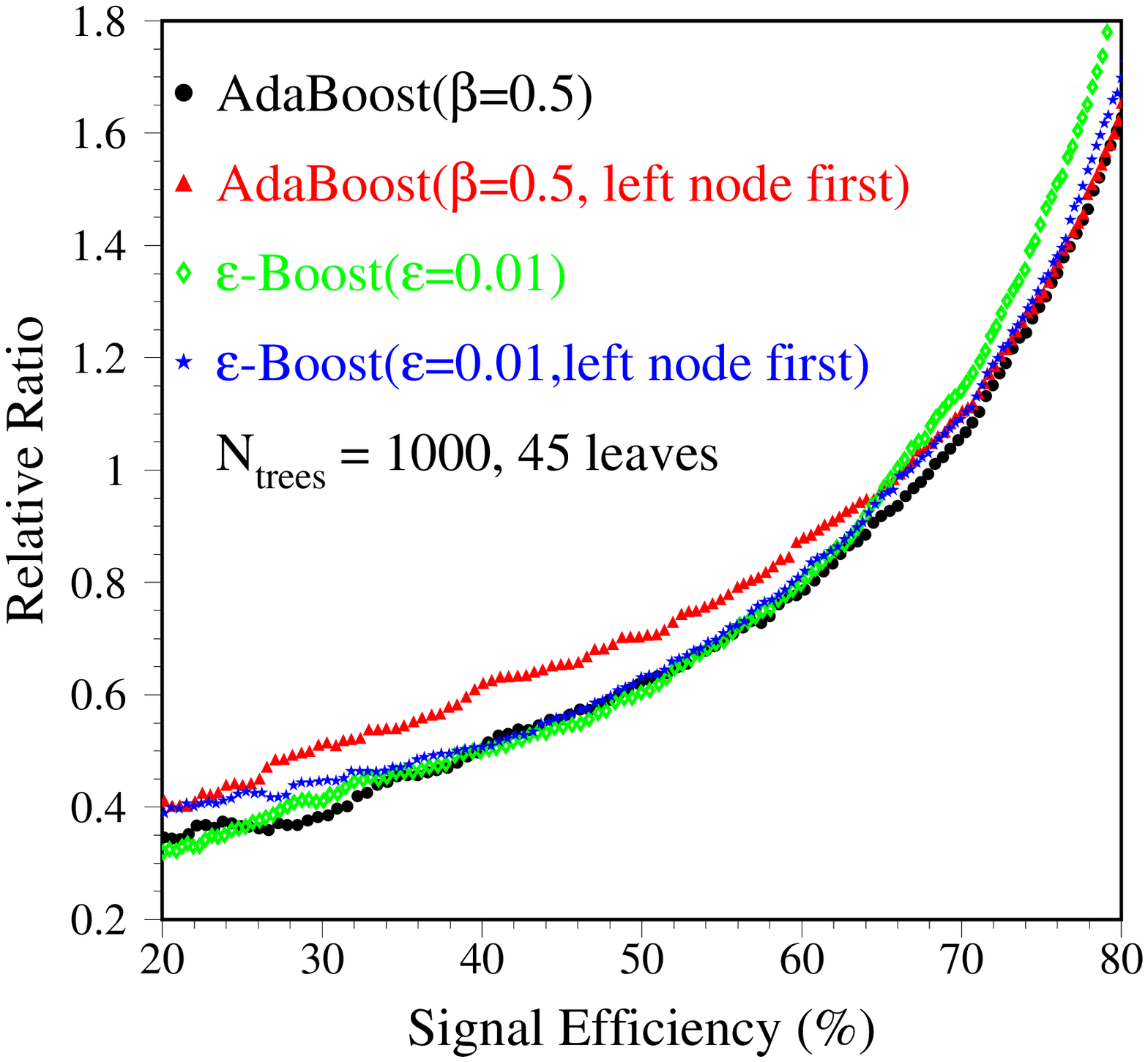,width=7.5cm}
\caption{Tuning $\epsilon$-Boost (left) and comparing two ways 
to split the tree branches (right).}
\end{figure}

\begin{figure}
\epsfig{figure=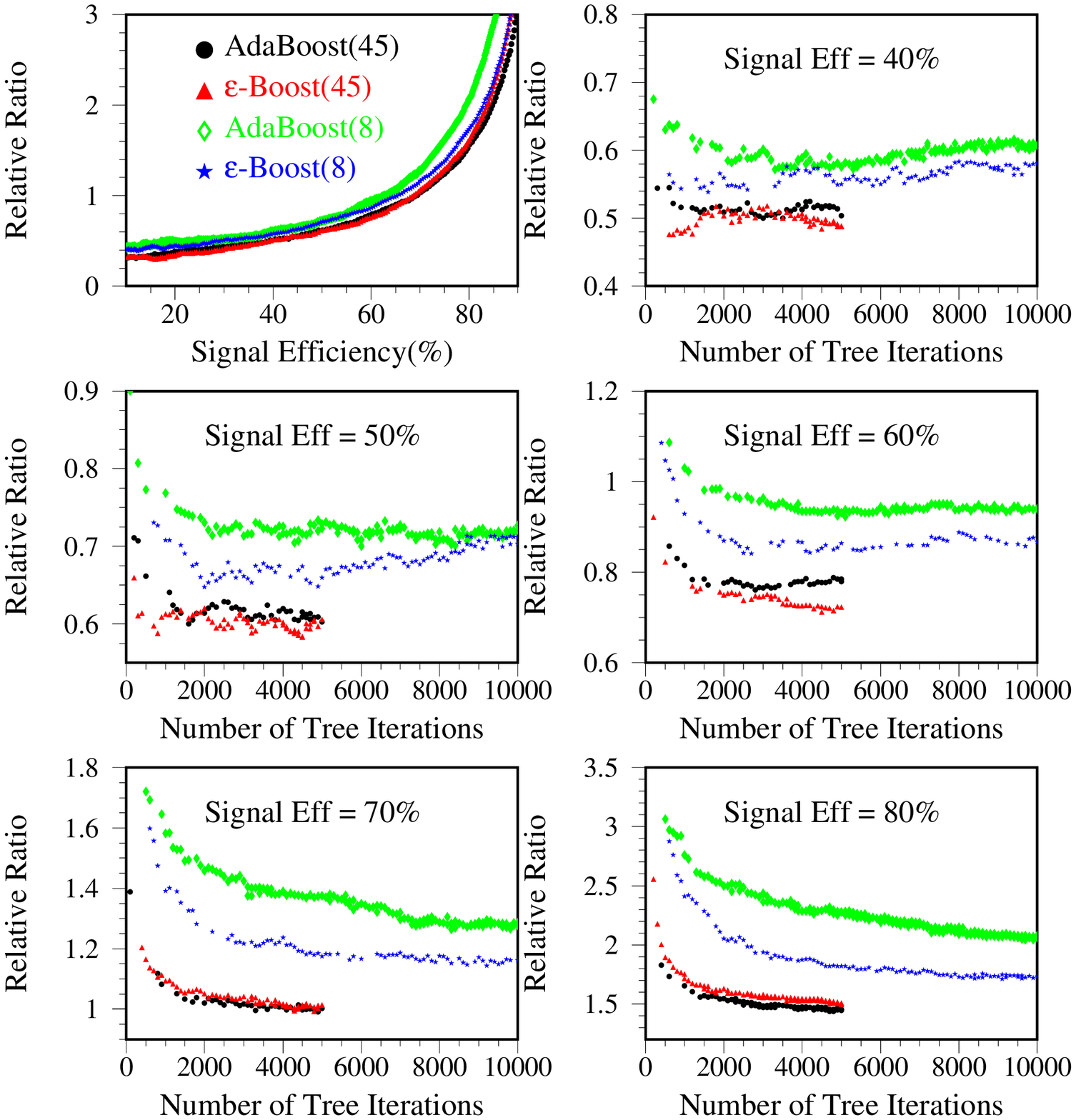,width=14cm}
\caption{Boostings with different tree sizes.}
\end{figure}

Larger $\epsilon$ makes $\epsilon$-Boost converge more quickly, but increasing the 
the size of decision trees also makes $\epsilon$-Boost converge more quickly. 
The performance comparison of AdaBoost with
different tree sizes shown in the right plot of Figure 2 is for the same number of
tree iterations (1000).
To make a fair comparison for the boosting performance with different tree sizes, 
it is better to let them have a similar number of total tree leaves.
The top left plot of the Figure 6 shows the relative ratio versus the signal 
efficiency for AdaBoost and $\epsilon$-Boost with similar numbers of the total
tree leaves, 1800 tree iterations for 45 leaves per tree and 10000 tree iterations
for 8 leaves per tree. 
For a small decision tree size of 8 leaves, the performance of  the $\epsilon$-Boost
is better than that of AdaBoost for 10000 trees. 
For a large decision tree size of 45 leaves, $\epsilon$-Boost 
has slightly better performance than AdaBoost at low $\nu_e$ signal 
efficiency ($<$65\%), but worse at high $\nu_e$ signal efficiency ($>$70\%). The 
comparison between small tree size (8 leaves) and large tree size (45 leaves) with 
comparable overall decision tree leaves indicates that large tree size with 45 leaves 
yields $\sim$10\%-20\% better performance for the MiniBooNE Monte Carlo samples.

The other five plots in Figure 6 show the relative ratio versus the number of 
tree iterations for AdaBoost and $\epsilon$-Boost with 45 leaves and 8 leaves assuming 
signal efficiencies of 40\%, 50\%, 60\% ,70\% ,80\%, respectively.
The maximum number of tree iterations is 5000 for the large tree size of 45 leaves and 10000
for the small tree size of 8 leaves. 
Usually, the performance of the boosting method 
becomes better with more tree iterations
in the beginning; then at some point, it may reach an optimal value and 
gradually get worse with increasing number of trees, especially in the low
signal efficiency region. The turning point of the boosting performance 
depends on the signal efficiency and MC samples used for training and test.

Generally, if the number of weighted signal events is larger than the number of 
weighted background events
in a given leaf, it is called a signal leaf, otherwise, a background leaf.
Here, the threshold value for signal purity is 50\%
for the leaf to be called a signal leaf.
This threshold value can be modified, say, to 30\%, 40\%, 45\%, 60\% or 70\%.
It is seen in Figure 7 that
the performance of boosted decision trees with Adaboost
degrades for threshold values away from the central value of 50\%.
Especially for
threshold values away from 50\%, the $err_m$ of m$th$ tree often converges
to 0.5 within about
100 tree iterations; after that the weights of the misclassified events do
not successfully update because
$\alpha_m \equiv \beta \times \ln((1-err_m)/err_m) = 0$ if $err_m=0.5$. Then
$w_i = w_i \times e^{\alpha_m \times I(y_i \ne T_m(x_i))}$
remains the same as for the previous
tree.  Typically, the
$err_m$ value increases for the first 100-200 tree iterations and then
remains stable for further tree iterations, causing the weight of m$th$ tree,
$\alpha_m$, to decrease for the first 100-200 tree iterations and then
remain stable.  For practical use of the AdaBoost algorithm,
a lower limit, say, 0.01, on $\alpha_m$ will avoid the
impotence of the succeeding boosted decision trees.

This problem is unlikely to happen for $\epsilon$-Boost because the weights
of misclassified events are always updated by the same factor,
$e^{2\epsilon \times I(y_i \ne T_m(x_i))}$.
If differing purity threshold values are applied
to boosted decision trees with $\epsilon$-Boost,
the performance  peaks around 50\% and slightly worsens,
typically within 5\%, for other values ranging from 30\% to 70\%.

\begin{figure}
\begin{center}
\epsfig{figure=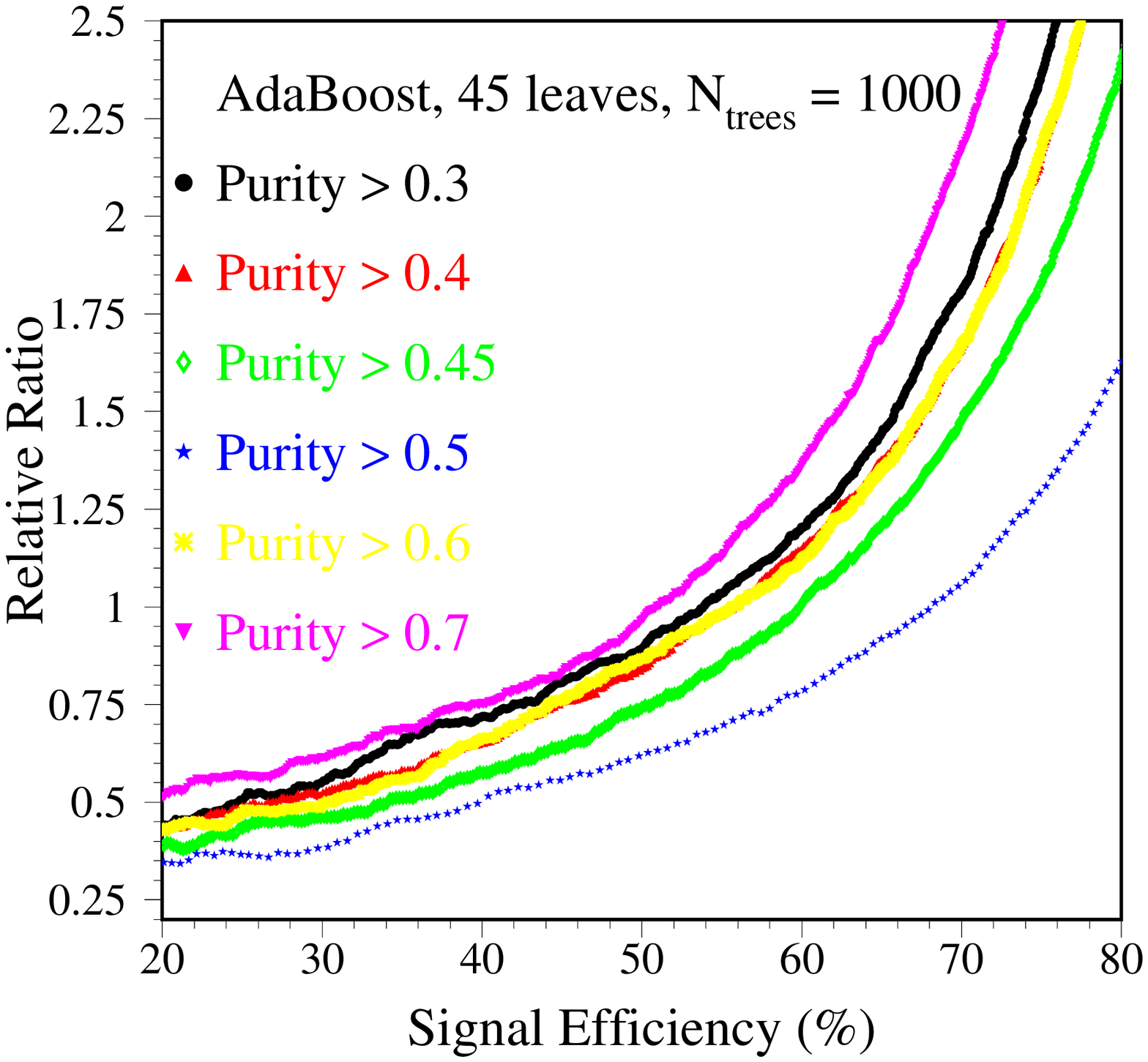,width=10cm}
\caption{Performance of AdaBoost with $\beta=0.5$, 45 leaves per tree and
1000 tree iterations for
various threshold values for signal purity to determine whether the tree leaf
is signal leaf or background leaf.}
\end{center}
\end{figure}

\begin{figure}
\begin{center}
\epsfig{figure=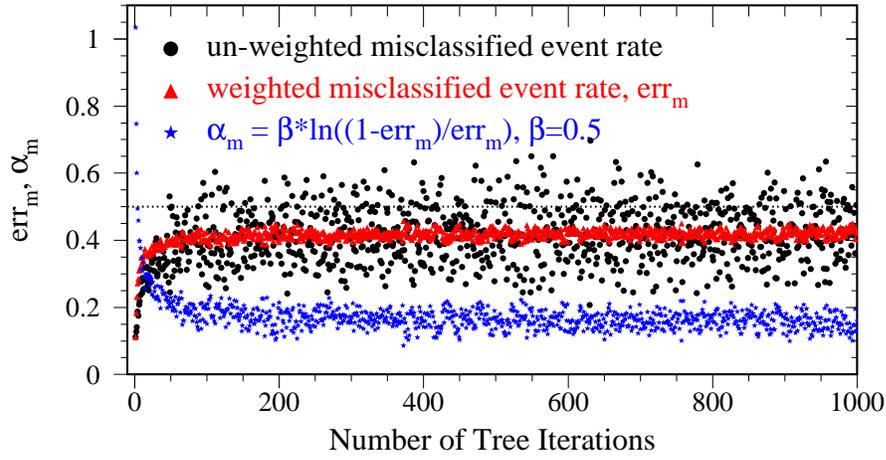,width=12cm}
\caption{The unweighted, weighted misclassified event rate ($err_m$), $\alpha_m$ 
versus the number of tree iterations for AdaBoost with $\beta = 0.5$ and signal purity
threshold value of 50\%.}
\end{center}
\end{figure}

The unweighted misclassified event rate, weighted
misclassified event rate $err_m$ and $\alpha_m$ for the boosted decision 
trees with the AdaBoost algorithm versus the number of tree iterations 
are shown in Figure 8,
for a signal purity threshold value of 50\%.
From this plot, it is clear that, after a few hundred tree iterations,
an individual boosted
decision tree  has a very weak discriminant
power (i.e., is a ``weak'' classifier).  The $err_m$ is about 0.4-0.45,
corresponding to $\alpha_m$ of around 0.2-0.1.  The unweighted event
discrimination of an individual tree is even worse, as is also seen in
Figure 8.
Boosted decision trees focus on the
misclassified events which usually have high weights after hundreds
of tree iterations. The advantage of the boosted decision trees is that the method
combines all decision trees, "weak" classifiers, to make a powerful 
classifier as stated in the Introduction section.

When the weights of misclassified events are increased (boosted),
some events which are very difficult correctly classify obtain
large event weights. In principle, some outliers which
have large event weights may degrade the boosting performance. To avoid
this effect, it might be useful to set an upper limit for the 
event weights to trim some outliers. It is found that setting a  
weight limit doesn't improve the boosting performance, and, in fact, may
degrade the boosting performance slightly.  However, 
the effect was observed to be within one standard
deviation for the statistical error. One might also trim
events with very low weights which can be correctly classified easily
to provide a better chance for difficult events. No apparent improvement or 
degradation was observed considering the statistical error.
These results may indicate that the boosted decision trees have the
ability to deal with outliers quite well and to focus on the events
located around the boundary regions where it is difficult to
correctly distinguish signal and background events.

\section{Comparison Among Boosting Algorithms}

Besides AdaBoost and $\epsilon$-Boost, there are other
algorithms such as $\epsilon$-LogitBoost, and $\epsilon$-HingeBoost which
use different ways of updating the event weights for the misclassified
events. The four plots of Figure 9 show the relative ratio versus
the signal efficiency for various boostings with different tree sizes.
The top left, top right, bottom left, and bottom right plots are for 
500, 1000, 2000, and 3000 tree iterations, respectively. 
Boosting with a large tree size of 45 leaves is seen to work better than 
boosting with a small tree size of 8 leaves as noted above. AdaBoost and
$\epsilon$-Boost have comparable performance, 
slightly better than that of $\epsilon$-LogitBoost. 
$\epsilon$-HingeBoost is the worst among these four boosting algorithms, 
especially for the low signal efficiency region. 

\begin{figure}
\epsfig{figure=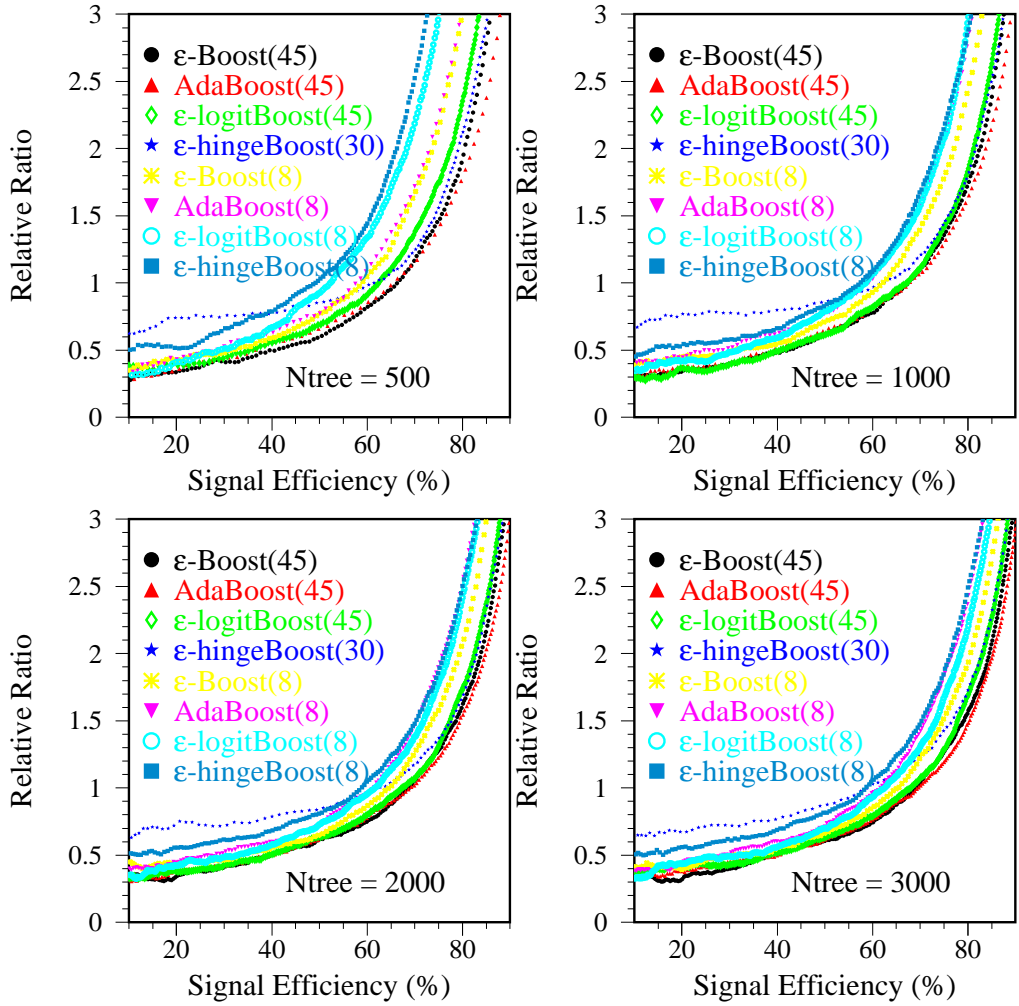,width=14cm}
\caption{Performance comparison of various Boostings.}
\end{figure}

\begin{figure}
\epsfig{figure=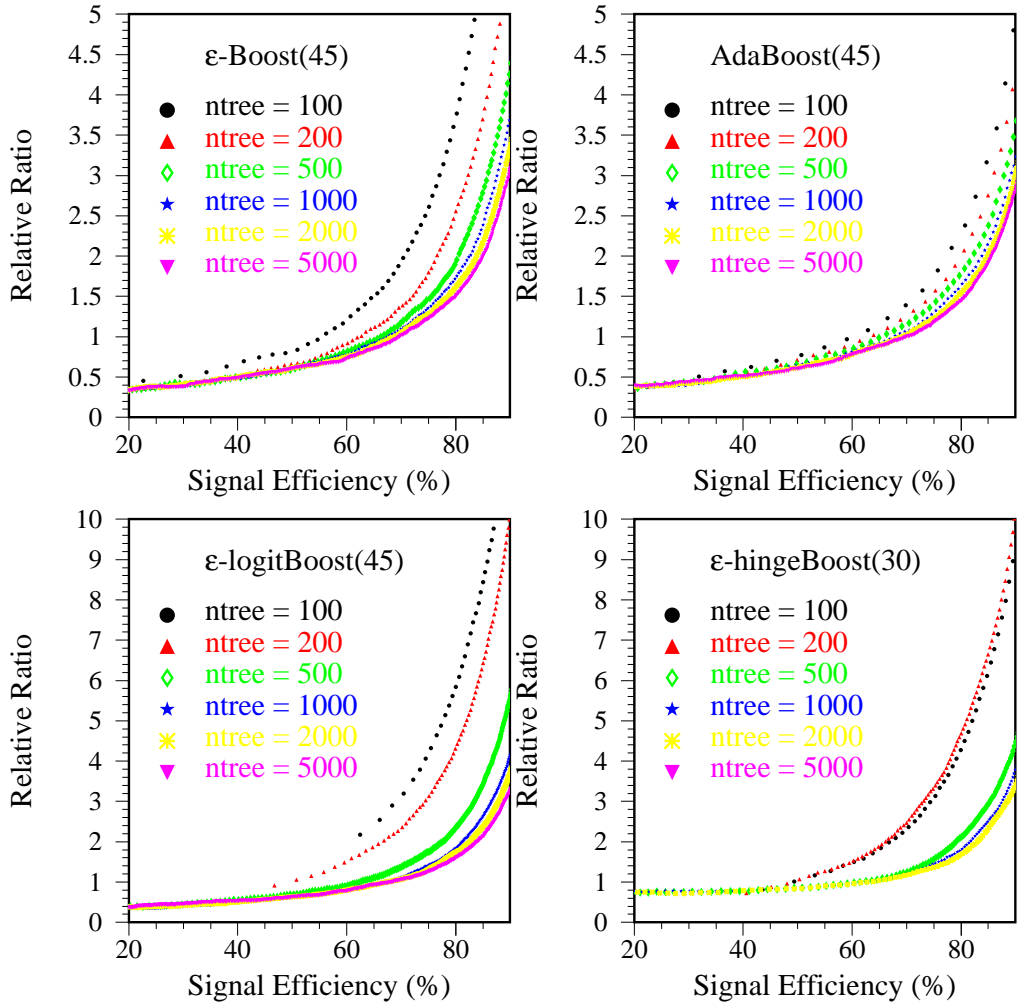,width=14cm}
\caption{Performance comparison of various Boostings with large tree size.}
\end{figure}

\begin{figure}
\epsfig{figure=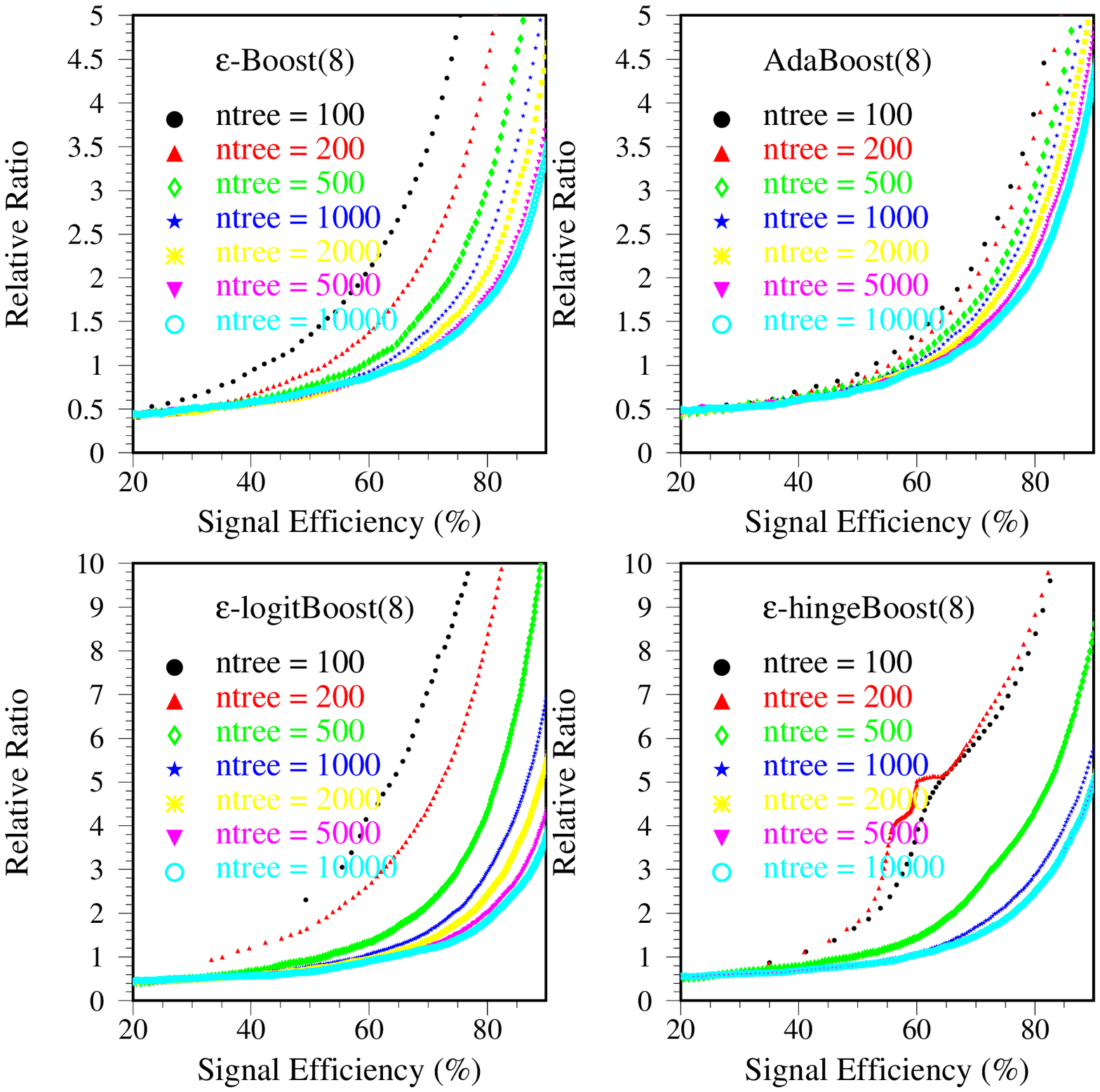,width=14cm}
\caption{Performance comparison of various Boostings with small 
tree size of 8 leaves per tree.}
\end{figure}

The top left, top right, bottom left and bottom right plots of Figure 10
show the relative ratio versus the signal efficiency with 45 leaves of 
$\epsilon$-Boost, AdaBoost, $\epsilon$-LogitBoost and $\epsilon$-HingeBoost
for varying numbers of tree iterations.
Generally, boosting performance continuously improves
with an increase in the number
of tree iterations until an optimum point is reached.
From the two top plots, it is apparent that $\epsilon$-Boost
converges more slowly than does AdaBoost; however, with about
1000 tree iterations, their performances are very comparable. There is only
marginal improvement beyond 1000 tree iterations for high
signal efficiency, and the performance may get worse for the low signal efficiency region
if the boosting is over-trained (goes beyond the optimal performance range).
Similar plots for the four boosting algorithms 
with 8 leaves per decision tree are shown 
in the Figure 11. Results for $\epsilon$-HingeBoost with 30 and 8 tree leaves are shown in 
the bottom right plots of Figures 10 and 11.  The performance for 200 tree iterations
seems worse than that for 100 tree iterations. This may indicate that its performance is 
unstable in the first few hundred tree iterations, but works well after about 500 tree
iterations. However, the overall performance of $\epsilon$-Hinge boost
is the worst among the four boosting algorithms
described above.

For some purposes, LogitBoost has been found to be superior to other 
algorithms\cite{dettling2003}.  For the
MiniBooNE data, it was found to have about 10\%-20\% worse background
contamination for a fixed signal efficiency than the regular AdaBoost.
LogitBoost converged very rapidly after less than 200 trees and the
contamination ratio got worse past that point.  A modification of 
LogitBoost was tried in which the convergence was slowed by taking
$T(x) = \sum_{m=1}^M \epsilon \times {1\over 2} T_m(x)$, the extra factor
of $\epsilon = 0.1$ slowing the weighting update rate.  This indeed improved the
performance considerably, but the results were still slightly worse than
obtained with AdaBoost or $\epsilon$-Boost for a tree size of 45 leaves.  
The convergence to an optimum point still took fewer than 300 trees, 
which was less than the number needed with AdaBoost or $\epsilon$-Boost.

Gentle AdaBoost and Real AdaBoost were also tried; both of them were found 
slightly worse than the discrete AdaBoost.
Relative error ratio versus signal efficiency for various boosting algorithms
are listed in Table.\ref{table:ratio}.

\begin{table}
\begin{center}
\begin{tabular}{|c|c|c|c|c|c|c|c|} \hline\hline
Boosting   & Parameters & \multicolumn{6}{|c|}{Relative ratios for given signal efficiencies} \\ \hline
Algorithms & $\beta$,$\epsilon$ ($N_{leaves},N_{trees}$) &  ~30\%~ & ~40\%~ & ~50\%~ & ~60\%~ & ~70\%~ & ~80\%~ \\ \hline\hline
AdaBoost  & 0.3 (45,1000) & 0.39 & 0.49 & 0.63 & 0.80 & 1.12 & 1.73 \\ \hline
AdaBoost  & 0.5 (45,1000) & 0.38 & 0.50 & 0.62 & 0.78 & 1.06 & 1.63 \\ \hline
AdaBoost  & 0.8 (45,1000) & 0.45 & 0.54 & 0.62 & 0.82 & 1.07 & 1.60 \\ \hline
AdaBoost  & 1.0 (45,1000) & 0.48 & 0.55 & 0.67 & 0.81 & 1.07 & 1.60 \\ \hline\hline
AdaBoost  & 0.5 (8,1000)  & 0.53 & 0.63 & 0.81 & 1.11 & 1.78 & 3.21 \\ \hline
AdaBoost  & 0.5 (20,1000) & 0.43 & 0.58 & 0.71 & 0.93 & 1.31 & 2.20 \\ \hline
AdaBoost  & 0.5 (45,1000) & 0.38 & 0.50 & 0.62 & 0.78 & 1.06 & 1.63 \\ \hline
AdaBoost  & 0.5 (100,1000)& 0.43 & 0.51 & 0.61 & 0.76 & 1.00 & 1.45 \\ \hline\hline

$\epsilon$-Boost & 0.005 (45,1000)& 0.38 & 0.47 & 0.62 & 0.84 & 1.26 & 2.23 \\ \hline
$\epsilon$-Boost & 0.01 (45,1000) & 0.41 & 0.50 & 0.60 & 0.80 & 1.14 & 1.87 \\ \hline
$\epsilon$-Boost & 0.02 (45,1000) & 0.40 & 0.48 & 0.62 & 0.77 & 1.08 & 1.71 \\ \hline
$\epsilon$-Boost & 0.03 (45,1000) & 0.38 & 0.48 & 0.58 & 0.75 & 1.03 & 1.62 \\ \hline
$\epsilon$-Boost & 0.04 (45,1000) & 0.40 & 0.50 & 0.60 & 0.75 & 1.02 & 1.57 \\ \hline
$\epsilon$-Boost & 0.05 (45,1000) & 0.40 & 0.47 & 0.60 & 0.79 & 1.07 & 1.61 \\ \hline\hline

AdaBoost (b=0.5) & 0.5 (45,1000) & 0.39 & 0.47 & 0.60 & 0.76 & 1.06 & 1.58 \\ \hline
$\epsilon$-Boost (b=0.5) & 0.01 (45,1000) & 0.36 & 0.46 & 0.62 & 0.83 & 1.23 & 2.00 \\ \hline
$\epsilon$-Boost (b=0.5) & 0.03 (45,1000) & 0.38 & 0.45 & 0.58 & 0.76 & 1.06 & 1.65 \\ \hline
$\epsilon$-Boost (b=0.5) & 0.05 (45,1000) & 0.37 & 0.44 & 0.58 & 0.74 & 1.03 & 1.58 \\ \hline\hline

AdaBoost  & 0.5 (8,1000)  & 0.53 & 0.63 & 0.81 & 1.11 & 1.78 & 3.21 \\ \hline
AdaBoost  & 0.5 (8,5000)  & 0.50 & 0.60 & 0.74 & 0.98 & 1.40 & 2.52 \\ \hline
$\epsilon$-Boost & 0.01 (8,1000) & 0.49 & 0.55 & 0.71 & 0.93 & 1.40 & 2.44 \\ \hline
$\epsilon$-Boost & 0.01 (8,5000) & 0.51 & 0.55 & 0.66 & 0.86 & 1.17 & 1.82 \\ \hline
$\epsilon$-LogitBoost & 0.01 (8,1000) & 0.49 & 0.59 & 0.79 & 1.07 & 1.58 & 2.95 \\ \hline
$\epsilon$-LogitBoost & 0.01 (8,5000) & 0.52 & 0.57 & 0.68 & 0.89 & 1.22 & 2.01 \\ \hline
$\epsilon$-HingeBoost & 0.01 (8,1000) & 0.58 & 0.66 & 0.83 & 1.09 & 1.68 & 2.88 \\ \hline
$\epsilon$-HingeBoost & 0.01 (8,5000) & 0.61 & 0.69 & 0.82 & 1.05 & 1.48 & 2.49 \\ \hline\hline

$\epsilon$-LogitBoost & 0.01 (45,1000) & 0.39 & 0.50 & 0.61 & 0.82 & 1.11 & 1.84 \\ \hline
$\epsilon$-HingeBoost & 0.01 (30,1000) & 0.77 & 0.80 & 0.86 & 0.96 & 1.20 & 1.80 \\ \hline
LogitBoost & 1.0 (45,130) & 0.41 & 0.55 & 0.73 & 0.98 & 1.43 & 2.40 \\ \hline
LogitBoost & 0.1 (45,150) & 0.44 & 0.52 & 0.62 & 0.82 & 1.23 & 2.00 \\ \hline
Real AdaBoost & (45,1000)  & 0.47 & 0.57 & 0.69 & 0.82 & 1.10 & 1.60 \\ \hline
Gentle AdaBoost & (45,1000)& 0.47 & 0.54 & 0.67 & 0.83 & 1.05 & 1.56 \\ \hline\hline

Random Forests(RF) & (400,1000)             & 0.49 & 0.63 & 0.85 & 1.29 & 1.92 & 3.50 \\ \hline
AdaBoosted RF & 0.5 (100,1000) & 0.48 & 0.56 & 0.66 & 0.81 & 1.04 & 1.58 \\ \hline
\end{tabular}
\end{center}
\caption{\label{table:ratio} Relative error ratio versus signal efficiency for 
various boosting algorithms for MiniBooNE data. Differences up to about
0.03 are largely statistical. b=0.5 means smooth scoring function described in Section 9.}
\end{table}

\section{Comparison of AdaBoost and Random Forests}

The random forests is another algorithm which uses a "majority vote" to
improve the stability of the decision trees.
The training events are selected randomly with or without
replacement. Typically,  one half or one third of the training
events are selected for each decision tree training.
The input variables can also be selected randomly for determining 
the tree splitters. There is no event weight update for the
misclassified events. For the AdaBoost algorithm, each tree is
built using the results of the previous tree; for the random 
forests algorithm, each tree is independent of the other trees.
\begin{figure}
\begin{center}
\epsfig{figure=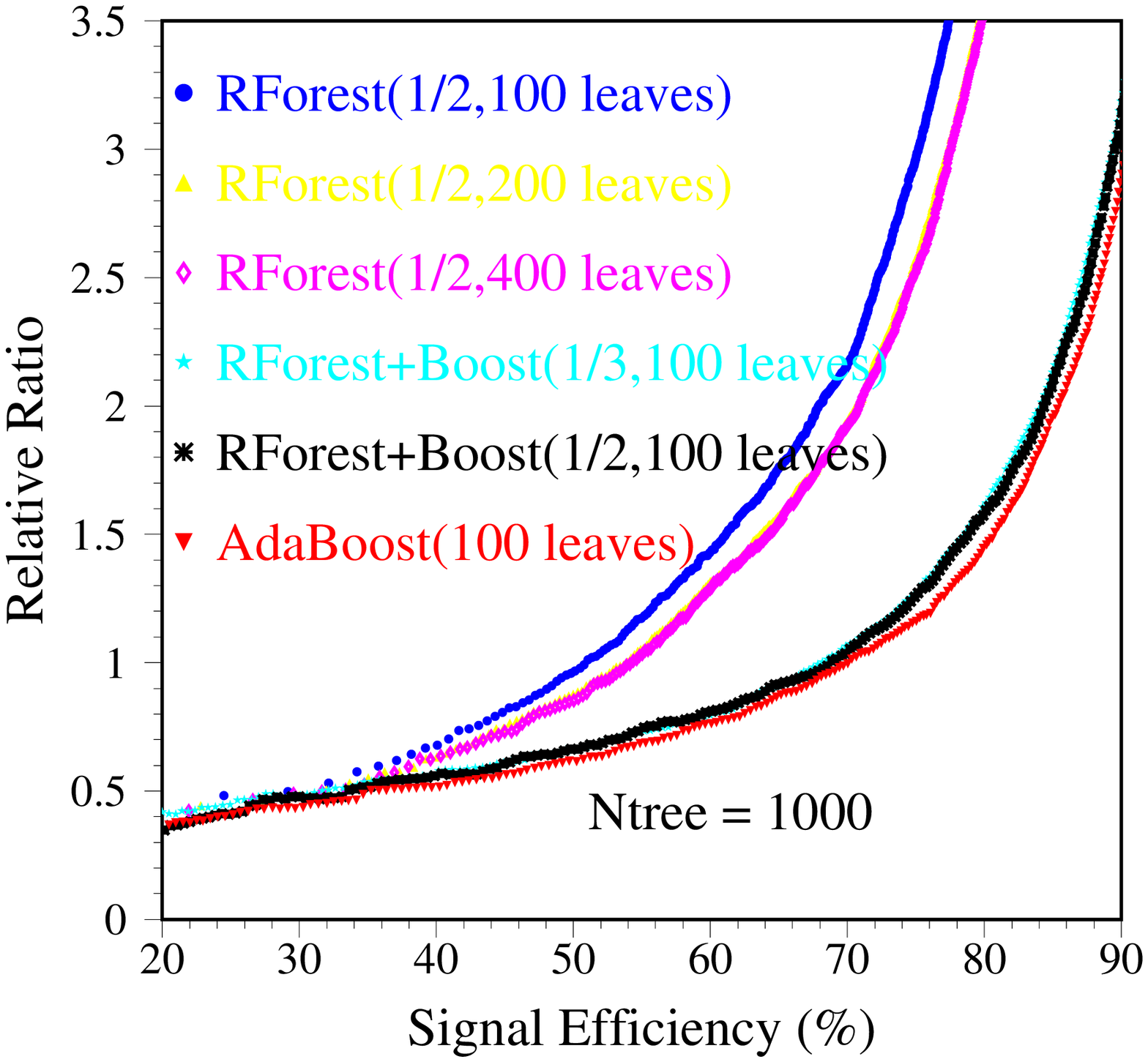,width=10cm}
\caption{Performance comparison of AdaBoost, random forests and
boosted random forests with 1000 tree iterations.}
\end{center}
\end{figure}

Figure 12 shows a comparison between random forests of different 
tree sizes and Adaboost, both with 1000 tree iterations.
Large tree size is preferred
for the random forests (The original random forests method lets
each tree develop fully until
all tree leaves are pure signal or background).  In this study a
fixed number of tree leaves were used.
The performance of the random forests
algorithm with 200 or 400 tree leaves is about equal.
Compared with
AdaBoost, the performance of the random forests method is significantly
worse. The main reason for the inefficiency is that there is no
event weight update for the misclassified events. One of main advantages
for the boosting algorithm is that the weights of misclassified events are
boosted which makes it possible for them to be correctly classified in 
succeeding tree iterations. 

Considering this advantage, an 
event weight update algorithm (AdaBoost) was used to boost the random forests.
The performances of the boosted random forests algorithm are then significantly better
than those of the original random forests as can be seen in Figure 12.
The performance of the AdaBoost with 100 leaves per decision tree is 
slightly better than that of the boosted random forests.
Other tests were made using one half training events selected randomly
for each tree together
with 30\%, 50\%, 80\% or 100\% of the
input variables selected randomly for each
tree split.  The performances of the boosted random forests method using the
AdaBoost algorithm are very stable.

The boosted random forests only uses one half or one third of the
training events selected randomly for each tree and also
only a fraction of the input variables for each tree split,
selected randomly; This method has the advantage that it can run
faster than regular AdaBoost while providing similar performance.
In addition, it may also help to avoid over-training since the training
events are selected partly and randomly for each decision tree. 

\section{Post-fitting of the Boosted Decision Trees}

Some recent papers \cite{zhou2002,friedman2003,friedman2004} indicate that post-fitting of 
the trained boosted decision trees may help to make further improvement. 
One possibility is that a selected ensemble of many decision trees could be better than 
the ensemble of all trees.
Here post-fitting of the weights of decision trees was tried.  The basic idea
is to optimize the boosting performance by retuning the weights of the decision trees or
even removing some of them by setting them to have 0 weight. 
A Genetic Algorithm \cite{GA75,GA88} is used to 
optimize the weights of all trained decision trees. 

A new MC sample is used for this purpose. The MC sample is split into three subsamples, 
mc1, mc2 and mc3, each subsample having about 26700 signal events and 21000 background events.
Mc1 is used to train AdaBoost with 1000 decision trees.  The background 
efficiency for mc1, mc2 and mc3 for a signal efficiency of 60\% are 0.12\%, 5.15\% and
4.94\%, respectively. If mc1 is used for post-fitting, 
then the corresponding background efficiency
can be driven down to 0.05\%, 
but the background efficiency for test sample mc3 is about 5.5\%. It has
become worse after post-fitting. It seems that it is not good to 
use same sample for the boosting
training and post-fitting. If mc2 is used for post-fitting, then the background efficiency
goes down to 4.21\% for the mc2, and 4.76\% for the testing sample mc3. The relative improvement
is about 3.6\% and the statistical error for the background events is about 3.2\%.
Suppose the MC samples for post-fitting and testing are exchanged, 
mc3 is used for post-fitting while mc2 is used for
testing. The background efficiency is 4.38\% for training sample mc3 and 5.06\% for the testing sample
mc2. The relative improvement is about 1.5\%. 

A second post-fitting program was tried, the Pathseeker program of J.H. Friedman and
B.E. Popescu\cite{friedman2003,friedman2004},
a robust regularized linear regression and classification method.
This program produced no overall improvement, with perhaps a marginal 4\%
improvement for 50\% signal efficiency.
It seems that post-fitting makes only a marginal improvement
based on our studies.

\section{How to Select Input Variables}

\begin{figure}
\epsfig{figure=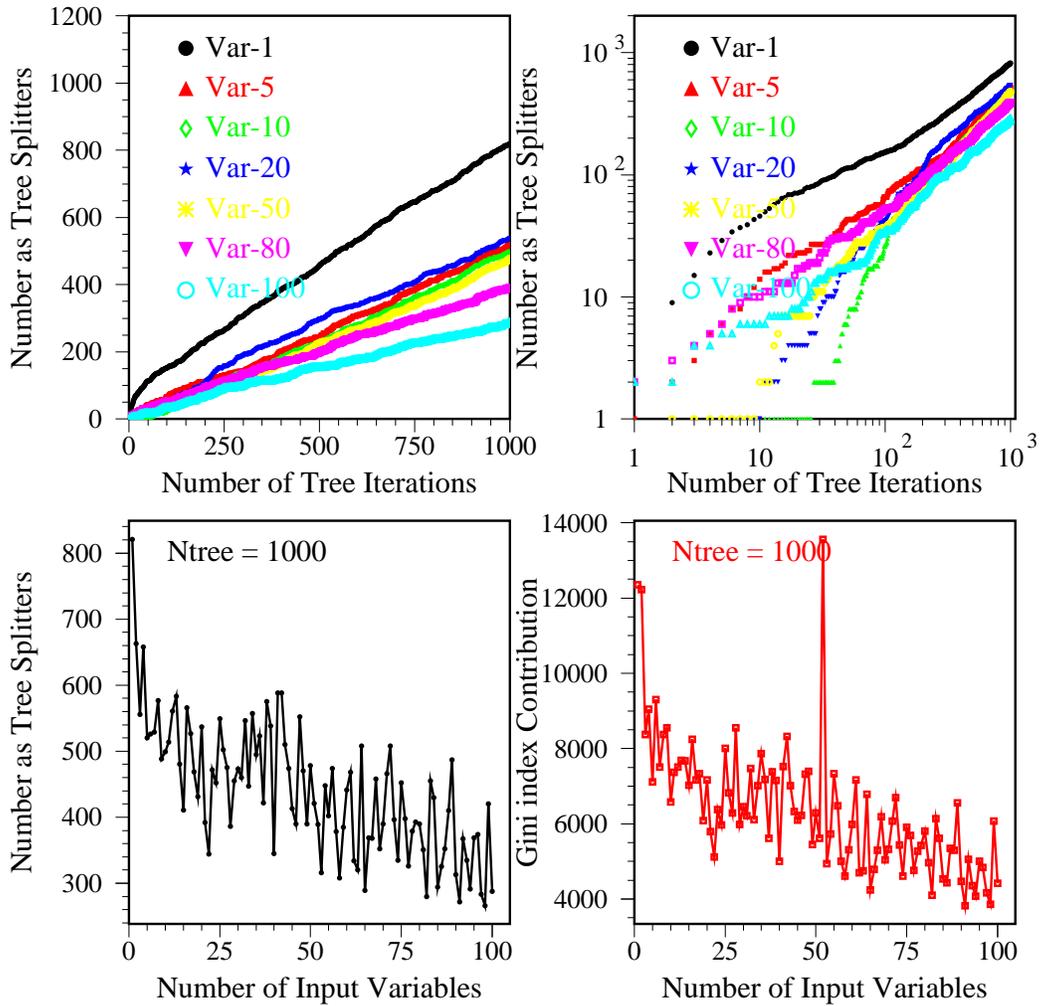,width=14cm}
\caption{Top plots show the dependence of the number of times a variable is 
used as a tree splitter versus the number tree iterations for some selected
input variables. The top left plot is linear scale and the top right
plot is log scale.
The number of times a variable is used as a tree splitters versus all 
input variables with 1000 tree iterations
is shown in the bottom left plot.  The bottom right plot shows
the corresponding Gini index contribution for all input variables.}
\end{figure}

One of the major advantages of the boosted decision tree algorithm 
is that it can handle large numbers of input variables as was 
pointed out previously\cite{nima-boosting2005}. 
Generally speaking, more input variables cover more
information which may help to improve signal and background event separation.
Often one can reconstruct several hundreds or even thousands of variables which 
have some discriminant power to separate signal and background events. Some of 
them are superior to others, and some variables may have correlations with others. 
Too many variables, some of which are ``noise'' variables, won't improve but may
degrade the boosting performance. It is useful to select the most useful
variables for boosting training to maximize the performance. New MC samples
were generated with 182 reconstructed variables. In order to select the most
powerful variables, all 182 variables were used as input to boosted decision
trees running 150 tree iterations. Then the effectiveness of the input variables
was rated based on how many times each variable was used as a tree splitter. The first
variable in the sorted list was regarded as the most useful variable for  
boosting training. The first 100 sorted input variables were selected to train 
AdaBoost with $\beta=0.5$, 45 leaves per decision tree and 1000 tree iterations.
The dependence of the number of times a variable is used as a tree splitter
versus the number of tree iterations is shown for
some selected input variables (Variables number 1, 5, 10, 20, 50, 80 and 100)
in the top left plot with
linear scale and in the top right plot with log scale.

In this way, the first 30, 40, 60, 80, 100, 120 
and 140 input variables were selected from the sorted list to train boosted decision trees
with 1000 tree iterations. A comparison of their performance is shown in the
left plot of Figure 14. The boosting performance steadily improves with
more input variables until about 100 to 120. 
Adding further input variables doesn't improve and may degrade
the boosting performance. The main reason for the degradation is that there is
no further useful information in the additional input variables and these variables can 
be treated as ``noise'' variables for the boosting training. However, if the
additional variables include some new information which is not included in the
other variables, they should help to improve the boosting performance. 

So far only one way to sort the input variables has been described. 
Some other ways can also be used and work reasonably well as shown in the right
plot of Figure 14. List1 means the input variables are sorted based
on the how many times they were used as tree splitters for 150 tree iterations,
list2 means the input variables are sorted based on their Gini index
contributions for 150 tree iterations, and list3 means the variables are 
sorted according to which variables are
used earlier than others as tree splitters for 150 tree iterations. List1001,
list1002 and list1003 are similar to list1, list2 and list3,
but use 1000 tree iterations. The first 100 input variables from the sorted lists
are used for boosting training with 1000 tree iterations. The performances
are comparable for 100 input variables sorted in different ways. However, the 
boosting performances for list1 and list3 are slightly better than the others. 

If an equal number of input variables of 100 are selected from each list, the number
of variables which overlap typically varies from about 70 to 90 for the different lists. 
In other words, about 10 to 30 input variables are different among the various lists.
In spite of these differences, the boosting performances are still comparable and stable. 
Further studies with MC samples generated using varied MC input parameters 
corresponding to systematic errors show that
the boosting outputs are very stable even though some input variables vary quite a lot.
If these same varied MC samples are applied to the ANNs, it turns out that 
boosted decision trees work significantly better than the ANNs 
for both event separation performance and for stability.

\begin{figure}
\epsfig{figure=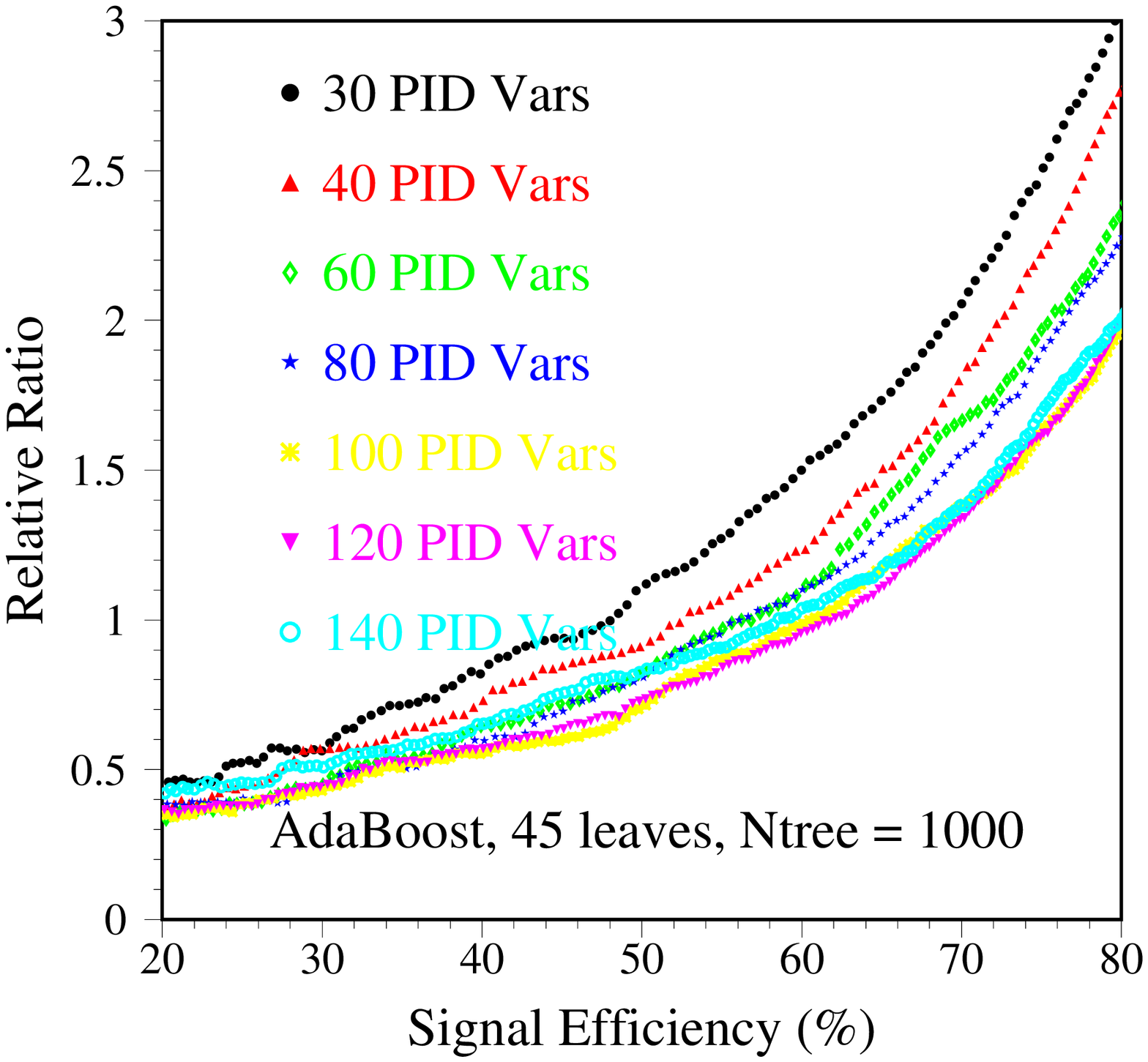,width=7.5cm}
\epsfig{figure=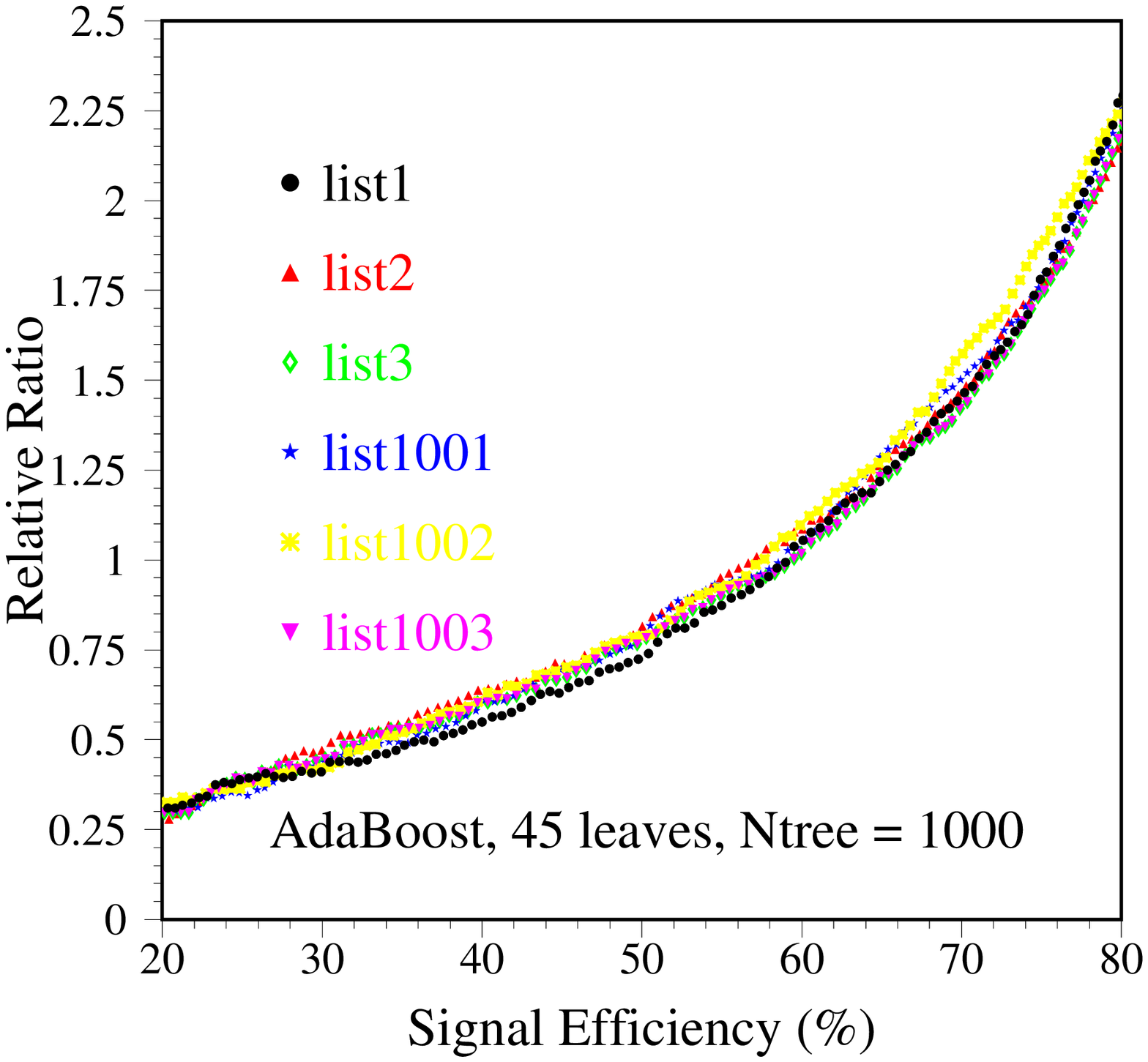,width=7.5cm}
\caption{Left: boosting performances with different number of input variables.
Right: boosting performances with the 100 input variables selected
by using different ways. Totally, 1000 tree iterations are used.}
\end{figure}

\section{Tests of the Scoring Function}

In the standard boost, the score for an event from an individual tree is
a simple square wave depending on the purity of the leaf on which the event lands.
If the purity is greater than 0.5, the score is 1 and otherwise it is $-1$.
\begin{figure}
\begin{center}
\epsfig{figure=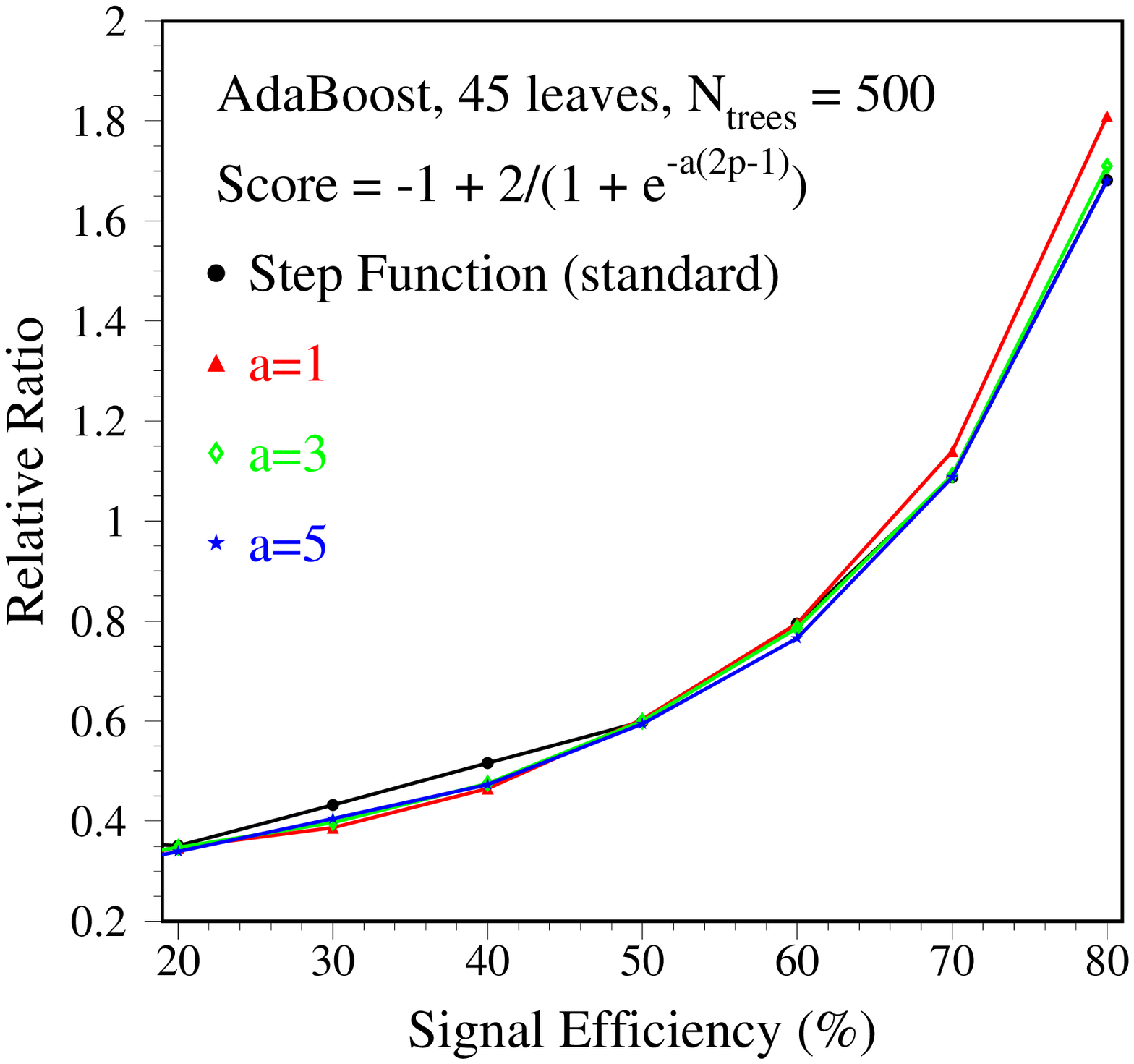,width=7.5cm}
\epsfig{figure=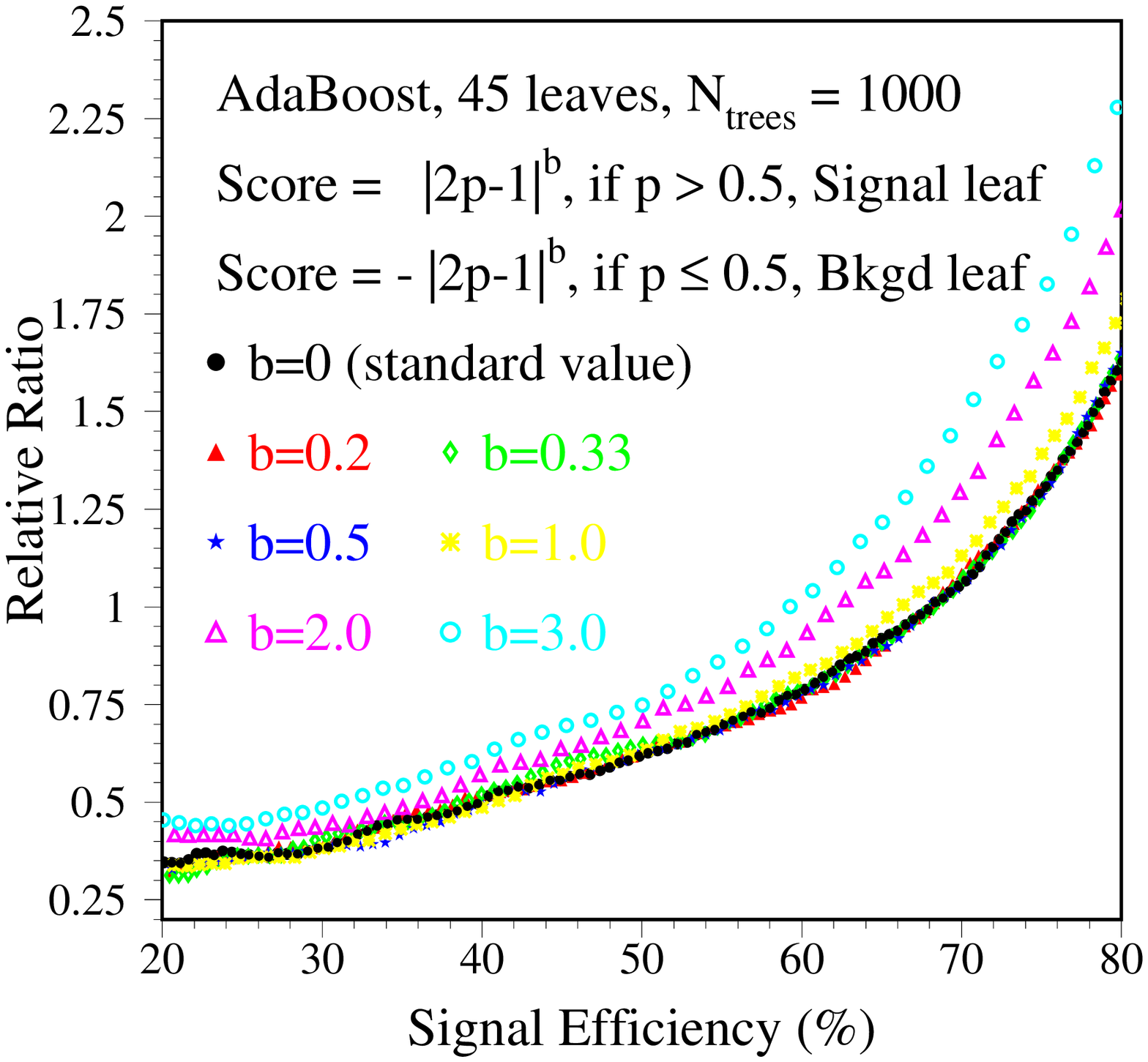,width=7.5cm}
\caption{Performance of AdaBoost with $\beta=0.5$, 
45 leaves per tree, and purity threshold value 0.5
for various parameters $a$ (left) and $b$ (right) values.}
\end{center}
\end{figure}

One can ask whether a smoother function of the purity
might be more appropriate.  If the purity of a leaf is 0.51, should the
score be the same as if the purity were 0.99?  Two possible alternative
scores were tested.  Let $z = 2\times$purity$-1$.
$$\ \ \ A.\ \ {\rm score}= -1 + {2\over e^{-az}+1},$$
$$\ \ \ B.\ \ {\rm score}= {\rm sign}(z)\times |z|^b,$$
where $a$ and $b$ are parameters.

Tests were run for various parameter values for scores A and B and compared
with the standard step function.
Performance comparisons of AdaBoost for various parameters $a$ (left) and 
$b$ (right) values are shown in Figure 15. 

For a smooth function with $b=0.5$, boosting performance converges
faster than the original AdaBoost algorithm for the first few hundred 
decision trees, as shown in Figure 16. However, no evidence was found
that the optimum was reached any sooner by the smooth function.
 The reason is that the smooth function 
of the purity describes the probability of a given event to be signal or
background in more detail than the step function used in the original
AdaBoost algorithm. With an increase in the number of tree iterations, however,
the ``majority vote'' plays the most important role for the event separation.
The ultimate performance of the smooth function with $b=0.5$ 
is comparable to the performance of the standard AdaBoost.

\begin{figure}
\begin{center}
\epsfig{figure=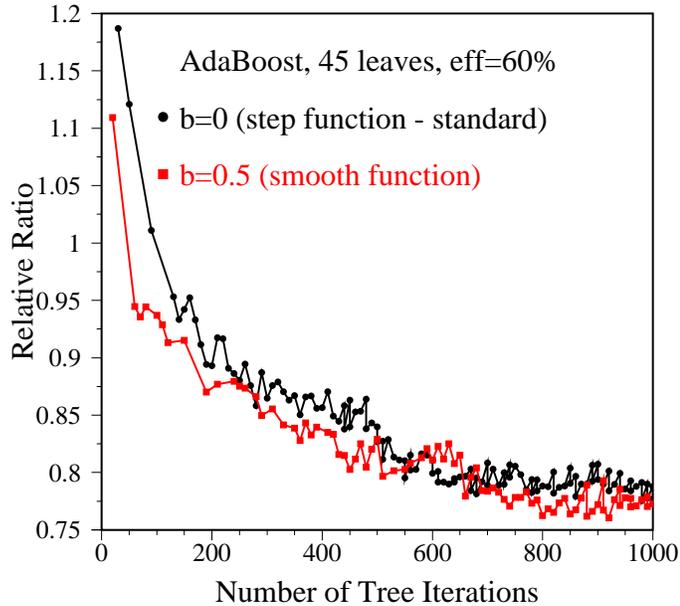,width=9cm}
\caption{Performance of AdaBoost with $b=0$ (step function) and
$b=0.5$ (smooth function), $\beta=0.5$,  45 leaves per tree, 
versus tree iterations.}
\end{center}
\end{figure}

\section{Some miscellaneous tests}

In MiniBooNE, one is trying to improve the signal to background ratio by
more than a factor of 100.  One might expect that one should start by
giving the background a greater total weight than the signal.  In fact,
giving the background two to five times the weight of the signal
slightly degraded the performance.  Giving the background 0.5 to 0.2 of the
weight of the signal gave the same performance as equal initial weights.

For one set of Monte Carlo runs the PID variables were carefully modified to
be flat as functions of the energy of the event and the event location within the
detector.  This decreased the correlations between the PID variables.
The performance of these corrected variables was compared with
the performance of the uncorrected variables.  As expected, the convergence
was much faster at first for the corrected variable boost.  However, as the
number of trees increased, the performance of the uncorrected variable boost
caught up with the other.  For 1000 trees, the performance of the two
boost tests was about the same.  Over the long run, boost is able to
compensate for correlations and dependencies, but the number of trees
for convergence can be considerably shortened by making the PID variables
independent.

The number of MC events used to train the boosting effectively
is also an important issue we have investigated. Generally, more training
events are preferred, but it is impractical to generate unlimited MC
events for training. The performance of AdaBoost with 1000 tree iterations,
45 tree leaves per tree using various number of background events ranging
from 10000 to 60000 for training are shown in Figure 17, where the number
of signal events is fixed 20000. For the MiniBooNE data, 
the use of
30000 or more background events works 
fairly well; fewer background events 
for training degrades the performance.

\begin{figure}
\begin{center}
\epsfig{figure=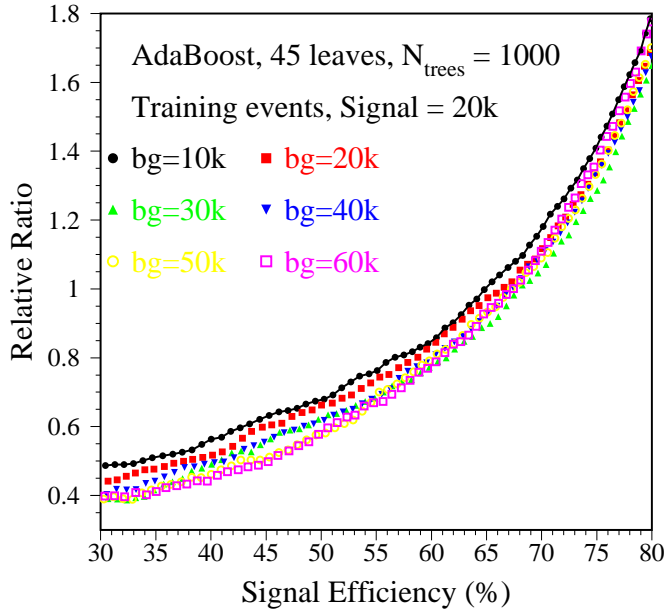,width=9cm}
\caption{Performance of AdaBoost with $\beta=0.5$, 
45 leaves per tree and 1000 tree iterations, 
using various number of background events for training.
20,000 signal events were used for the training.}
\end{center}
\end{figure}

\section{Conclusions}

PID input variables obtained using the event reconstruction programs 
for the MiniBooNE experiment were used to train boosted decision trees for 
signal and background event separation.
Numerous trials were made to tune the boosted decision trees. Based on the performance
comparison of various algorithms, decision trees with the AdaBoost or the $\epsilon$-Boost 
algorithms are superior to the others. The major advantages of boosted decision trees
include their stability, their ability to handle large number of input variables, 
and their use of boosted weights for misclassified events to give these events 
a better chance to be correctly classified in succeeding trees.

Boosting is a rugged classification method.  If one provides sufficient 
training variables and sufficient leaves for the tree, it appears that
it will, eventually, converge to close to an optimum value.  This assumes that
$\epsilon$ for $\epsilon$-Boost or $\beta$ for Adaboost are not set too \
large.  
There are modifications of the basic boosting procedure 
which can
speed up the convergence.  Use of a smooth scoring function improves
initial convergence.  In the last 
section, it was seen that removing
correlations of the input PID variables improved convergence speed.  For some
applications, the use of a boosted natural forests technique may also
speed the convergence.

For a large set of discriminant variables, several techniques can be used to 
select 
a set of powerful input variables to use for training boosted decision trees. 
Post-fitting of the boosted decision trees makes 
only a marginal improvement in the tests presented here.

\section{Acknowledgments}

We wish to express our gratitude to the MiniBooNE collaboration for the excellent
work on the Monte Carlo simulation and the software package for physics analysis.
This work is supported by the Department of Energy and the 
National Science Foundation of the United States.




\end{document}